\documentclass{JHEP3}
\usepackage{amssymb}
\usepackage{amsmath}
\usepackage{graphicx}
\newtheorem{theorem}{Theorem}

\newtheorem{proposition}[theorem]{Proposition}

\def\func#1{\mathop{\rm #1}\nolimits}

\title{Triangulated Surfaces in Twistor Space: a Kinematical Set up for Open/Closed String Duality}

\author{Mauro Carfora, Claudio Dappiaggi, Valeria L. Gili \\Dipartimento di Fisica Nucleare e Teorica, Universit\`{a} degli Studi di Pavia, \\
via A. Bassi 6, I-27100 Pavia, Italy, \\
and\\
Istituto Nazionale di Fisica Nucleare, Sezione di Pavia, \\
via A. Bassi 6, I-27100 Pavia, Italy\\ E-mail: \email{mauro.carfora@pv.infn.it}, \email{claudio.dappiaggi@pv.infn.it}, \email{valeria.gili@pv.infn.it}}

\abstract{
We exploit the properties of the hyperbolic  space $\mathbb{H}^{3}$ to discuss a simplicial setting for open/closed string duality based on (random) Regge triangulations decorated with null twistorial fields. We explicitly show that the twistorial N-points function, describing Dirichlet correlations  over the moduli space of open N-bordered genus g surfaces, is naturally mapped into the Witten-Kontsevich intersection theory over the moduli space of N-pointed closed Riemann surfaces of the same genus. We also discuss various aspects of the geometrical setting which connects this model to $PSL(2,\mathbb{C})$ Chern-Simons theory.
}

\preprint{hep-th/0607146}

\keywords{String dualities, simplicial quantum
gravity, hyperbolic geometry}
\begin{document}

\section{\protect\bigskip Introduction}
Major advances \cite{Gop-Vafa, Gop1, Gop2, Gopproc, Gop3, Gop4, Gai-Ras, Aharony, Akhmedov} in our understanding of Open/Closed string duality have provided a number of paradigmatical connections   between Riemann moduli space theory, piecewise-linear geometry, and the study of the gauge/gravity correspondence. These connections have a two-fold origin.  On the mathematical side they are deeply related to the fact that  moduli space admits  natural (semi-simplicial) decompositions which are in a one-to-one correspondence with classes of suitably decorated graphs. On the physical side they are consequence of the observation that these very decorated (Feynman) graphs  parametrize consistently the quantum dynamics of conformal and gauge fields. In a rather general sense, simplicial techniques provide a natural kinematical framework within which we can discuss open/closed string duality.
A basic problem in such a setting is to provide an explanation of how  open/closed duality is dynamically generated. In particular how a closed surface is related to a corresponding open surface, with gauge-decorated boundaries, in such a way that the quantization of such a correspondence leads to a open/closed duality. Typically, the natural candidate for such a mapping is Strebel's theorem which allows to  reconstruct a closed N-pointed Riemann surfaces $M$ of genus g out of the datum of a the quadratic differential associated with a ribbon graph \cite{strebel, mulase} . Ribbon graphs are open Riemann surfaces which one closes by inserting punctured discs, (so generating semi-infinite cylindrical ends). The dynamics of gauge fields decorating the boundaries of the ribbon graph is naturally framed within the context of boundary conformal field theory (BCFT) which indeed plays an essential role in the onset of a open/closed duality regime. The reason for such a relevance is to be seen in the fact that BCFT is based on algebraic structures parametrized by the  moduli
space of genus g Riemann surfaces with N  punctures $\mathcal{M}(g;N)$. This parametrization is deeply connected  with Strebel's theorem in the sense that it is consistent with the operation of sewing together any two ribbon graphs (open surfaces) with (gauge-decorated) boundaries, provided that we match the complex structure and the decoration in the overlap and keep track of which puncture is ingoing and which is outgoing.  In such a setting a BCFT leads to a natural algebra, over the decorated cell decomposition of Riemann moduli space, which can be related to the algebra of physical space of states of the theory and to their boundary dynamics. It is fair to say that in such a sense BCFT realizes open/closed duality as the quantization of a gauge-decorated Strebel's mapping.

\medskip

\noindent It is well-known that in the analysis of the cellular geometry of Riemann moduli space there is also another point of view, pionereed by R. Penner and W. Thurston (see \emph{e.g.} \cite{Penner1, ThurstonB}), not emphasizing the role of conformal geometry, but rather exploiting the parametrization of the moduli space in terms of hyperbolic surfaces. Here, one deals with hyperbolic surfaces with punctures (\emph{i.e.}, surfaces with cuspidal geometry) rather than with surfaces with marked points. Moreover, in such a setting one generates a combinatorial decomposition of Riemann moduli space, still parametrized by ribbon graphs, not by using quadratic differentials but rather via the geometry of surface geodesics. Some geometrical aspects of the role of this particular combinatorial parametrization in open/closed string duality has been recently discussed by R. Kaufmann and R. Penner \cite{Kaufmann}. One of the advantages of the hyperbolic point of view is that one has a clear picture of the geometry of the surface. In particular of the reasons why, in assembling a surface out of ideal hyperbolic triangles, one may get an open surface with boundary, (even if the glueing pattern of the triangulation is combinatorially consistent with a closed surface) \cite{ThurstonB}. This gives a geometrical mechanism describing the transition between closed and open surfaces which, in a dynamical sense, is more interesting than Strebel's construction. The drawback is that in such a setting BCFT is not readily available and it is not obvious, (at least to us), how to formulate open/closed string duality, by, so to say, quantizing the dynamics of gauge fields on such  hyperbolic  combinatorial decomposition of moduli space.  

\medskip

The purpose of this paper is to discuss a bridge between these two combinatorial formalisms which seems to select the best of the two approaches. We use both Euclidean simplicial complexes (dual to the standard ribbon graphs) and hyperbolic geometry. This is made possible by exploiting the known correspondence existing  between locally Euclidean structures in dimension two and hyperbolic geometry in dimension three. Such a correspondence allows to define in a very suggestive way a map between closed surfaces $M$, $\partial M=\emptyset$, triangulated with Euclidean triangles, and open hyperbolic surfaces $\Omega $, $\partial \Omega \not=\emptyset$, triangulated by ideal hyperbolic triangles. The triangulated Euclidean closed surfaces we use are (random) Regge triangulations $|T_{l}|\rightarrow M$ with non-trivial curvature degrees of freedom, (represented by conical angles $\{\Theta (k)\}$, supported at its vertices, describing 2D-gravity), and decorated with null twistorial fields which can be thought of as defining the embedding of the triangulation in hyperbolic three-space $\mathbb{H}^{3}$. The hyperbolic surface $\Omega $ associated with such conical Euclidean triangulation  is generated by locally projecting the Euclidean triangles of $|T_{l}|\rightarrow M$ into the 2-dimensional boundary at infinity of $\mathbb{H}^{3}$. By considering the upper half-space model
$\mathbb{H}^{3,+}_{up}$ of $\mathbb{H}^{3}$, such a projection has an elementary realization in terms of the geometry of ideal tetrahedra whose vertices are decorated with (small) horospheres $\Sigma _{k}$. Any such a horosphere has an intrinsic Euclidean structure and can act as a projection screen $\Sigma _{\infty }$ from which Euclidean triangles can be mapped, via hyperbolic geodesics in $\mathbb{H}^{3}$, into ideal hyperbolic triangles with vertices decorated by horocycles. The geometry of this geodesic projection and of the induced horocyclic decoration is quite non-trivial. The vertices $\sigma ^{0}(k)$ of $|T_{l}|\rightarrow M$, with a conical defect $\Theta (k)$, get mapped into a corresponding geodesic boundary $\partial \Omega(k) $  of $\Omega $,  with a length given by $|\ln\frac{\Theta (k)}{2\pi }|$. Moreover, such boundaries come naturally endowed with  a SU(2) holonomy generated by a flat su(2) connection on $\Omega $. Null twistors naturally enter into this pictures as the consequence of the parametrization of the horospheres of $\mathbb{H}^{3}$ in terms of null vectors in 4-dimensional Minkowski space, (equivalently in terms of the twistorial description of the geodesics of $\mathbb{H}^{3}$).   
This correspondence between closed (singular Euclidean) surfaces and open hyperbolic surface is easily promoted to the corresponding moduli spaces: $\mathcal{M}_{g,N}\times \mathbb{R}_{+}^{N}$ the moduli spaces of $N$-pointed closed Riemann surfaces of genus g whose marked points are decorated with the given set of conical angles, and  $\mathcal{M}_{g,N}(L)$ the moduli spaces of open Riemann surfaces of genus g with $N$ geodesic boundaries decorated by the corresponding lengths. This provides a nice kinematical set up for establishing a open/closed string duality once the appropriate field decoration is activated. The simplest case is when we consider non-dynamical null twistors fields decorating the vertex of the Regge triangulation. These fields geometrically describe geodesics in $\mathbb{H}^{3}$, with an end point at 
$\infty\in \partial \mathbb{H}^{3,+}_{up}$ $=$ $(\mathbb{R}^{2}\times \{0\})\cup\{\infty \}$,  projecting to the $N$ components $\partial \Omega _{k}$ of the  boundary of $\Omega $. Thus, they can naturally be interpreted as fields on  $\Omega $ with preassigned Dirichlet boundary conditions on the $\partial \Omega _{k}$'s. At the level of the moduli space $\mathcal{M}_{g,N}(L)$ we can consider the N-point function on $\mathcal{M}_{g,N}(L)$, describing correlations between such Dirichlet conditions. By exploiting a remarkable result recently obtained by Maryam Mirzakhani \cite{mirzakhani1, mirzakhani2}, we can easily show that the such a correlation function  is naturally mapped into the generating function of the Witten-Kontsevich intersection theory on $\mathcal{M}_{g,N}\times \mathbb{R}_{+}^{N}$. We can also consider the decoration of $\Omega $ associated with the $\mathfrak{su}(2)$ flat connection naturally defined on $\Omega$ (again generated by the twistorial fields, since $SU(2)$ appears as the point stabilizer of $PSL(2,\mathbb{C})$ and one views $\mathbb{H}^{3}$ as the coset space $PSL(2,\mathbb{C})\setminus SU(2)$). In such a case one has a dynamic  $SU(2)$ Yang-Mills field defined on  $\Omega $ and it is straightforward to explicitly write down the corresponding $N$-points function on $\mathcal{M}_{g,N}(L)$. The analysis of  open/closed string duality in such a case is much more delicate since it also involves intersection theory over
the variety $Hom(\pi _{1}(\Omega ),SU(2))/SU(2)$
of representations, up to conjugacy, of the fundamental group of the 
bordered surface $\Omega $ in $SU(2)$. Such an intersection theory is 
related to a careful treatment of the corresponding BCFT on $\Omega $ 
and calls into play  Chern-Simons theory for the $PSL(2,\mathbb{C})$ 
group. This is still work in progress and it will be presented in a 
companion paper. However we thought appropriate, in a analysis mainly 
dealing with geometrical kinematics of open/closed duality, to conclude
our presentation by discussing the aspects of the mapping  between 
closed Regge triangulated surfaces $|T_{l}|\rightarrow M$ and open 
hyperbolic surfaces $\Omega $ which naturally activates  
$PSL(2,\mathbb{C})$ Chern-Simons theory. In simple terms the mechanism 
is just the parametrization of ideal tetrahedra in  $\mathbb{H}^{3}$ in 
terms of the similarity structure associated with Euclidean triangles. 
This allows to associate with the Regge triangulated surface 
$|T_{l}|\rightarrow M$ the hyperbolic volume of a three-dimensional 
ideal triangulation. When such a triangulation is actually the 
triangulation of a hyperbolic three-manifold then the celebrated 
Kashaev-Murakami volume conjecture \cite{Kashaev95, Kashaev97, Murakami, 
Hmurakami, CS} directly activates $PSL(2,\mathbb{C})$ Chern-Simons 
theory.

It is also interesting to remark that Mirzakhani's results have been
exploited by G. Mondello in dealing with the Poisson structure of
Teichmuller space of Riemann surfaces with boundaries \cite{Mondello}.
This can be of relevance in describing the symplectic structure of
$Hom\left(\pi_1(\Omega),SU(2)\right)/SU(2)$ on such bordered surfaces
and its interaction with $PSL(2,\mathbb{C})$ Chern-Simons theory.

 \medskip

\noindent \emph{Outline of the paper}. In section 2 we sketch the 
properties of metrically triangulated closed surfaces and of the 
associated locally Euclidean (singular) structures. This is a familiar 
subject in simplicial quantum gravity and non-critical string theory. 
Here we emphasize a few delicate aspects which are not so widely known 
and which are relevant in our setting. In section 3 we discuss the 
connection between triangulated surfaces, hyperbolic three-geometry and 
twistors. Such a connection is fully exploited in section 4 where we 
explicitly construct a mapping between closed metrically triangulated 
surfaces and hyperbolic surfaces with boundary. We extend such a mapping 
to the appropriate moduli spaces and discuss an explicit example of 
open/closed string duality. In section 5 we analyze the connection with 
hyperbolic three-geometry with emphasis on the relation between the 
computability of the (hyperbolic) three-volume in terms of the 
parameters of the original two-dimensional triangulation and 
$PSL(2,\mathbb{C})$ Chern-Simons theory. This latter section is not 
really instrumental to the main body of the paper, but we have 
nonetheless decided to include it to further show the deep connections 
existing between simplicial methods and the Physics of open/closed 
duality. As a final comment, we would like to add a disclaimer: since 
some of the geometrical techniques we exploit both in the simplicial 
formalism as well as in hyperbolic geometry may not be widely known, we 
have presented a rather detailed analysis rather than qualitative 
arguments. As is often the case with the serious use of simplicial 
techniques, the notation can become at a few points quite unwieldy, and 
for that we apologize to the reader.

\section{\protect\bigskip Random Regge triangulations }

 A proper understanding of the role that simplicial methods have in open/closed string duality requires that we consider metric triangulations of surfaces where both the connectivity and the edge-lengths of the triangulation are allowed to vary. Triangulations with fix connectivity but varying edge-length are known as Regge triangulations \cite{regge} wheras if we fix the edge-length and allow the connectivity to vary we get Dynamical triangulations \cite{carfora1, ambjorn}. Random Regge triangulations \cite{carfora}, (a term being something of an oxymoron), correspond to a geometrical realization whereby both connectivity and edge-length are allowed to fluctuate and provide the general framework of our analysis.

\medskip

\noindent Let $T$ denote an oriented finite $2$-dimensional
semi-simplicial complex  with underlying polyhedron $|T|$, {\emph i.e.}, a simplicial complex where the star 
$Star[\sigma ^{0}(j)] $ of a vertex $
\sigma ^{0}(j)\in T$ (the union of all triangles of which $\sigma ^{0}(j)$
is a face) is allowed to contain just one triangle. Denote respectively by $F(T)$, $E(T)$, and $V(T)$ the set of $N_{2}(T)$ faces, $N_{1}(T)$ edges, and $N_{0}(T)$ vertices of $T$, where $
N_{i}(T)\in \mathbb{N}$ is the number of $i$-dimensional subsimplices $\sigma
^{i}(...)$ $\in$ $T$. If we assume that $|T|$ is homeomorphic to a closed surface $M$ of genus $g$, then a
random Regge triangulation \cite{carfora} of $M$ is a realization of the homomorphism
$|T_{l}|\rightarrow {M}$ such that each edge 
$\sigma ^{1}(h,j)$ of $T$ is a rectilinear simplex of variable
length $l(h,j)$. In simpler terms, $T$ is generated by Euclidean triangles glued together by isometric identification of adjacent edges. It is important to stress that the connectivity of 
$T$ is not a priori fixed as in the case of standard Regge triangulations
(see \cite{carfora} for details). Henceforth, if not otherwise stated, when we speak of Regge surfaces we shall always mean a Random Regge triangulated surface. We also note that in such a general setting a (semi-simplicial)
dynamical triangulation $|T_{l=a}|\rightarrow {M}$ is a particular case \cite{ambjorn, carfora1} of a
random Regge PL-manifold realized by rectilinear and equilateral simplices
of a fixed edge-length $l=a$. 

\subsection{\protect\bigskip The metric geometry of triangulated surfaces}
\label{megeom}
The metric geometry of a random Regge triangulation is defined by the
distribution of edge-lengths $\sigma ^{1}(m,n)\rightarrow l(m,n)$ satisfying the appropriate triangle inequalities  
$l(h,j)\leq l(j,k)+l(k,h)$, whenever $\sigma ^{2}(k,h,j)\in F(T)$. Such an  assignment  uniquely characterizes the Euclidean geometry of the triangles $\sigma ^{2}(k,h,j)\in T$ and in particular, via the cosine law,
the associated vertex angles $\theta _{jkh}\doteq \angle \lbrack l(j,k),l(k,h)]
$,   $\theta _{khj}\doteq \angle \lbrack l(k,h),l(h,j)]$,  $\theta
_{hjk}\doteq \angle \lbrack l(h,j),l(j,k)]$; \emph{e.g.}
\begin{equation}
\cos \theta _{jkh}=\frac{l^{\,2}(j,k)+l^{\,2}(k,h)-l\,^{2}(h,j)}{2l(j,k)l(k,h)%
}\,.
\end{equation}
If we note that the area $\Delta (j,k,h)$ of $\sigma
^{2}(j,k,h)$ is provided, as a function  of $\theta
_{jkh}$, by
\begin{equation}
\Delta (j,k,h)=\frac{1}{2}l(j,k)l(k,h)\,\sin \,\theta _{jkh},
\label{areasin}
\end{equation}
then the angles $\theta _{jkh}$, $\theta _{khj}$, and $\theta
_{hjk}$ can be equivalently characterized by the formula
\begin{equation}
\cot \,\theta _{jkh}\,=\,\frac{l^{2}(j,k)+l^{2}(k,h)-l^{2}(h,j)}{4\,\Delta
(j,k,h)},
\end{equation}
which will be useful later on. It must be stressed that the assignment
\begin{gather}
\mathcal{E}(T):\left\{ \sigma ^{2}(k,h,j)\right\} _{F(T)}\longrightarrow 
\mathbb{R}_{+}^{3\,N_{2}(T)} \\
\sigma ^{2}(k,h,j)\longmapsto (\theta _{jkh},\theta _{khj},\theta _{hjk}) 
\notag
\end{gather}
of the angles $\theta _{jkh}$,  $\theta _{khj}$, and $\theta
_{hjk}$ (with the obvious constraints $\theta _{jkh}>0$, $\theta _{khj}>0$, $%
\theta _{hjk}>0$, and $\theta _{jkh}+\theta _{khj}+\theta _{hjk}=\pi $)   to each
$\sigma ^{2}(k,h,j)\in T$ does not allow to reconstruct the metric geometry of a Regge surface. $\mathcal{E}(T)$ only characterizes the local Euclidean structure (\cite{rivin1})  of
$|T_{l}|\rightarrow M$, \emph{i.e.} the similarity classes of the
realization of each  $\sigma ^{2}(k,h,j)$ as an Euclidean triangle; in simpler words, their shape and not their actual size.
As emphasized by
Rivin \cite{rivin1}, the knowledge of the locally Euclidean structure on $|T|\rightarrow M$
corresponds to the holonomy representation 
\begin{equation}
H(T):\pi _{1}(T\,\,\backslash \,V(T))\longrightarrow GL_{2}(\mathbb{R})
\end{equation}
of the fundamental group of the punctured surface $M\,\backslash \,V(T)$
into the general linear group $GL_{2}(\mathbb{R})$, and the action of $GL_{2}(\mathbb{R})$ is not rigid enough for defining a coherent Euclidean glueing of the corresponding triangles $\sigma
^{2}(k,h,j)\in T$. A few subtle properties of the geometry of Euclidean triangulations are at work here, and to put them to the fore  let us consider $q(k)$ triangles $\sigma ^{2}(k,h_{\alpha },h_{\alpha +1})$
incident on the generic vertex $\sigma ^{0}(k)\in T_{l}$ and generating the star
\begin{equation}
Star[\sigma ^{0}(k)]\doteq \cup _{\alpha =1}^{q(k)}\sigma ^{2}(k,h_{\alpha
},h_{\alpha +1}),\;\;h_{q(k)+1}\equiv h_{1}. 
\end{equation}
To any given  locally Euclidean structure
\begin{equation}
\mathcal{E}\left(Star[\sigma ^{0}(k)]\right)\doteq \left\{ (\theta _{\alpha+1
,k,\alpha},\,\theta _{k,\alpha,\alpha+1 },\,\theta _{\alpha,\alpha +1,k })\right\}
\end{equation}
 on $Star[\sigma ^{0}(k)]$\ \ there corresponds a conical defect  $\Theta (k)$ $\doteq $ $\sum_{\alpha =1}^{q(k)}\theta _{\alpha+1 ,k,\alpha}$ supported at $\sigma ^{0}(k)$, and a logarithmic dilatation \cite{rivin1}, with
respect to the vertex \ $\sigma ^{0}(k)$, of the generic triangle $\sigma
^{2}(k,h_{\alpha },h_{\alpha +1})$ $\in $ $Star[\sigma ^{0}(k)]$, \emph{i.e.,} 
\begin{equation}
D(k,h_{\alpha },h_{\alpha +1})\doteq \;\ln \,\sin \,\theta _{k,\alpha,\alpha +1}-\,\ln \,\sin \,\theta
_{\alpha,\alpha +1,k }\,.
\end{equation}
To justify this latter definition, note that if $\{l(m,n)\}$ is a distribution of edge-lengths to the triangles 
$\sigma ^{2}(k,h_{\alpha },h_{\alpha +1})$ of $Star[\sigma ^{0}(k)]$ compatible with $\mathcal{E}(T)$, then by identifying $\theta _{k,\alpha
,\alpha +1}$ with $\angle \lbrack l(k,h_{\alpha }),\,l(h_{\alpha },h_{\alpha +1})]
$ and $\theta _{\alpha,\alpha +1,k }$  with the angle corresponding to $\angle \lbrack \,l(h_{\alpha },h_{\alpha
+1}),\,l(h_{\alpha +1},k)]$, and by
exploiting the law of sines, we can equivalently write
\begin{equation}
D(k,h_{\alpha },h_{\alpha +1})= \;\ln \,\frac{l(h_{\alpha +1},k)}{%
l(k,h_{\alpha })} \,.
\end{equation}
In terms of this parameter, we can define (\cite{rivin1}) the dilatation holonomy of $Star[\sigma
^{0}(k)]$ according to
\begin{equation}
H(Star[\sigma ^{0}(k)])\doteq \sum_{\alpha =1}^{q(k)}\,D(k,h_{\alpha
},h_{\alpha +1}).
\end{equation}
The vanishing of $H(Star[\sigma ^{0}(k)])$
implies that if we circle around
the vertex $\sigma ^{0}(k)$, then the lengths $l(h_{\alpha +1},k)$ and $l(k,h_{\alpha +1})$ of \ the pairwise adjacent
oriented edges $\sigma ^{1}(h_{\alpha +1},k)$ and $\sigma ^{2}(k,h_{\alpha +1})$  match up for each $%
\alpha =1,...,q(k)$, with\ \ $h_{\alpha }=h_{\beta }$ if \ $\beta =\alpha \;%
\func{mod}\,q(k)$. A local Euclidean structure $\mathcal{E}(T)$  such that the dilatation holonomy $H(Star[\sigma ^{0}(k)])$ vanishes for each choice of star 
$Star[\sigma ^{0}(k)]$ $\subset T$ is called {\emph{conically complete}}. According to these remarks, the
triangles in  $T$ can be coherently glued into a random Regge triangulation,
with the  preassigned deficit  angles $\varepsilon (k)\doteq 2\pi -\Theta (k)$\ generated by
the given $\mathcal{E}(T)$, if and only if \ $\mathcal{E}(T)$ is complete in
the above sense. Note that if the deficit angles $\left\{ \varepsilon
(k)\right\} _{V(T)}$ \ all vanish, we end up in the more familiar notion of
holonomy associated with the completeness of the Euclidean structure
associated with $|T|\rightarrow M$ and described by a developing map whose
rotational holonomy around any vertex is trivial.

\subsection{\protect\bigskip Holonomies of singular Euclidean structures}

\noindent There is a natural way 
to keep track of both dilation factors and  conical defects by introducing a complex-valued holonomy.  \ Let $(\zeta
_{k},\zeta _{h},\zeta _{j})$ a ordered triple of complex numbers describing
the vertices of a realization, in the complex plane $\mathbb{C}$, of the
oriented triangle $\sigma ^{2}(k,h,j)$, with edge lenghts $l(k,h)$, $l(h,j)$%
, $l(j,k)$. By using Euclidean similarities we can always map \ $(\zeta
_{k},\zeta _{h},\zeta _{j})$ to $(0,1,\zeta _{_{jkh}})$, with
\begin{equation}
\zeta _{jkh}\doteq \frac{\zeta _{j}-\zeta _{k}}{\zeta _{h}-\zeta _{k}}=\frac{%
l(j,k)}{l(k,h)}\,e^{i\,\,\theta _{jkh}},
\end{equation}
where 
\begin{equation}
\arg \,\zeta _{jkh}=\arg \,(\zeta _{j}-\zeta _{k})-\arg \,(\zeta _{h}-\zeta
_{k})=\theta _{jkh}\in \lbrack 0,2\pi ),
\end{equation}
(thus $\func{Im}\,\zeta _{jkh}>0$).\ \ The triangle $(0,1,\zeta _{_{jkh}})$ is in the
same similarity class $\mathcal{E}(\sigma
^{2}(k,h,j))$ of \ $\sigma ^{2}(k,h,j)$, and the vector $\zeta
_{_{jkh}}$, the {\emph{complex modulus}} of the triangle $\sigma ^{2}(k,h,j)$ with respect to $%
\sigma ^{0}(k)$, parametrizes $\mathcal{E}(\sigma
^{2}(k,h,j))$.
The same similarity class is obtained by cyclically permuting the vertex which
is mapped to $0$, \emph{i.e.}, $(\zeta _{h},\zeta _{j},\zeta
_{k})\rightarrow (0,1,\zeta _{_{khj}})$ and $(\zeta _{j},\zeta _{k},\zeta
_{h})\rightarrow (0,1,\zeta _{_{hjk}})$, where 
\begin{equation}
\zeta _{khj}\doteq \frac{\zeta _{k}-\zeta _{h}}{\zeta _{j}-\zeta _{h}}=\frac{%
l(k,h)}{l(h,j)}\,e^{i\,\,\theta _{khj}},
\end{equation}
\begin{equation}
\zeta _{hjk}\doteq \frac{\zeta _{h}-\zeta _{j}}{\zeta _{k}-\zeta _{j}}=\frac{%
l(h,j)}{l(j,k)}\,e^{i\,\,\theta _{hjk}},
\end{equation}
are the moduli \ of $\mathcal{E}%
(\sigma ^{2}(k,h,j))$ with respect to the vertex $\sigma ^{0}(h)$ and $\sigma
^{0}(j)$, respectively. Elementary geometrical considerations imply that
the triangles  $(\zeta _{jkh}\,\,\zeta _{hjk}\,\,\zeta
_{khj},0,\zeta _{hjk}\,\zeta _{jkh})$, $(0,1,\zeta _{jkh})$, and $(\zeta _{jkh},\zeta
_{hjk}\,\zeta _{jkh},0)$ are congruent. This yields the relations 
\begin{gather}
\zeta _{jkh}\,\,\zeta _{hjk}\,\zeta _{khj}=-1, \\
\zeta _{jkh}\,\,\zeta _{hjk}=\zeta _{jkh}-1,  \notag
\end{gather}
according to which a choice of a moduli with respect a particular vertex
specifies also the remaining two moduli. For instance, if we describe $%
\mathcal{E}(\sigma ^{2}(k,h,j))$ by the modulus $\zeta _{jkh}\doteq \zeta $, ($\func{Im}\,\zeta >0$),
with respect to $\sigma ^{0}(k)$ then we get 
\begin{eqnarray}
\zeta _{jkh} &\doteq &\zeta , \\
\zeta _{khj} &=&\frac{1}{1-\zeta },  \notag \\
\zeta _{hjk} &=&1-\frac{1}{\zeta }.  \notag
\end{eqnarray}
By selecting the
standard branch on $\mathbb{C}-(-\infty ,0]$ of the natural logarithm, we also get
\begin{eqnarray}
\ln \zeta _{jkh} &=&\,\ln \zeta , \\
\ln \zeta _{khj} &=&-\ln \,(1-\zeta ),  \notag \\
\ln \zeta _{hjk} &=&\ln \,(1-\zeta )-\ln \,\zeta +\pi \,i.  \notag
\end{eqnarray}
In terms of these log-parameters we can extend the logarithmic
dilation of the generic
triangle $\sigma ^{2}(k,h_{\alpha },h_{\alpha +1})\in Star[\sigma ^{0}(k)]$,
to its complexified form 
\begin{equation}
D^{\mathbb{C}}(k,h_{\alpha },h_{\alpha +1})\doteq \ln \zeta _{h_{\alpha
+1},k,h_{\alpha }}=\ln \,\frac{l(h_{\alpha +1},k)}{l(k,h_{\alpha })}%
+i\,\,\theta _{h_{\alpha +1},k,h_{\alpha }},
\end{equation}
where $\zeta _{h_{\alpha +1},k,h_{\alpha }}$ is the complex modulus of the
triangle $\sigma ^{2}(k,h_{\alpha },h_{\alpha +1})$ with respect to the
vertex $\sigma ^{0}(k)$.\ \ Correspondingly we define
\begin{gather}
H^{\mathbb{C}}(Star[\sigma ^{0}(k)])\doteq \sum_{\alpha =1}^{q(k)}\,D^{%
\mathbb{C}}(k,h_{\alpha },h_{\alpha +1})= \\
=\sum_{\alpha =1}^{q(k)}\,D(k,h_{\alpha },h_{\alpha +1})+i\sum_{\alpha
=1}^{q(k)}\theta _{h_{\alpha +1},k,h_{\alpha }}=  \notag \\
=H(Star[\sigma ^{0}(k)])+i\,\,\Theta (k),  \notag
\end{gather}
where $\Theta (k)$ is the conical defect supported at the vertex $\sigma
^{0}(k)$. According to the previous remarks, it follows that the triangulation $%
|T_{l}|\rightarrow M$ will be conicaly complete iff \ $\func{Re}\,H^{\mathbb{C}%
}(Star[\sigma ^{0}(k)])=0$ for every vertex star, and its conical defects are
provided by $\left\{ \func{Im}\,H^{\mathbb{C}}(Star[\sigma ^{0}(k)])\right\} 
$. \ In
other words, a necessary and sufficient condition on the locally Euclidean
structure $\left\{ \theta _{jkh},\theta _{khj},\theta _{hjk}\right\} _{F(T)}$
in order to define a glueing and hence a random Regge triangulation is the
requirement that 
\begin{equation}
\prod_{k=1}^{q(k)}\zeta _{h_{\alpha +1},k,h_{\alpha }}\in U(1),
\end{equation}
for each  $Star[\sigma ^{0}(k)]$,  {\emph{i.e.}} that the image of $H^{\mathbb{C}}(T)$ lies in the group $U(1)$.
Note that the condition for having a flat Regge triangulation is stronger than $\prod_{k=1}^{q(k)}\zeta _{h_{\alpha +1},k,h_{\alpha }}=1$, $\forall k=1,..., N_{0}(T)$, since it requires that
\begin{equation}
\sum_{k=1}^{N_{0}(T)}H^{\mathbb{C}}(Star[\sigma ^{0}(k)])=2N_{0}(T)\pi \,i.
\end{equation}

\section{\protect\bigskip Regge surfaces and  ideal tetrahedra in $\mathbb{H}^{\,3}$}
\label{EHstructures}

The connection between similarity classes of arrangements of  Euclidean triangles with trivial holonomy $H^{\mathbb{C}}$ and triangulations of three-manifolds by ideal tetrahedra  is a well-known property of three-dimensional hyperbolic geometry \cite{benedetti}. This interplay extends in a subtle way  to the case in which $H^{\mathbb{C}}$ is no longer trivial (\emph{i.e.}, to  singular Euclidean structure \cite{rivin1}) and plays a key role in our results. 
\bigskip

\noindent To set the stage, let $\mathbb{H}^{3}$ denote the 3-dimensional hyperbolic space thought of as the subspace of Minkowski
spacetime $(M^{4},\langle \cdot ,\cdot \rangle )$ defined by 
\begin{equation}
\mathbb{H}^{3}\bigskip =\left\{ \overrightarrow{x}\doteq \left(
x^{0},x^{1},x^{2},x^{3}\right) \,\,|\;\left\langle \overrightarrow{x},%
\overrightarrow{x}\right\rangle =-1,\;x^{0}>0\right\} ,
\end{equation}
and equipped with the induced Riemannian metric defined by the restriction to
the tangent spaces $T_{x}\mathbb{H}^{3}$ 
of the\ standard Minkowski inner product 
\begin{equation}
\left\langle \overrightarrow{x},\overrightarrow{y}\right\rangle \doteq
-x^{0}y^{0}+x^{1}y^{1}+x^{2}y^{2}+x^{3}y^{3}.
\end{equation}
Recall that the group of orientation preserving isometries of $\mathbb{H}^{3}$ can be identified with the group 
$PSL(2,\mathbb{C})$ which acts transitively on $\mathbb{H}^{3}$ with point stabilizer provided by $SU(2)$.
Let $x\in\mathbb{H}^{3}$ and $\overrightarrow{y}\in T_{x}\mathbb{H}^{3}$ with 
$\left\langle \overrightarrow{y},%
\overrightarrow{y}\right\rangle =1$, then the geodesic in $\mathbb{H}^{3}$ starting at $x$ with velocity $\overrightarrow{y}$ is traced by the
intersection of $\mathbb{H}^{3}$ with the two-dimensional hyperplane of $M^{4}$ generated by the position vector $\overrightarrow{x}$ and the velocity $\overrightarrow{y}$ and is described by the mapping
\begin{equation}
\mathbb{R}\ni t\longmapsto \gamma (t)=\cosh (t)\,\overrightarrow{x}  + \sinh (t)\,\overrightarrow{y}.
\end{equation}
 Let $\gamma (\infty )$ denote the endpoint of $\gamma $ on the sphere at infinity $\partial \mathbb{H}^{3}\simeq \mathbb{S}^{2}$, a
closed horosphere centered at $\gamma (\infty )$ is a closed surface $\Sigma \subset 
\mathbb{H}^{3}$ which is orthogonal to all geodesic lines in $\mathbb{H}^{3}$
with endpoint $\gamma (\infty )$.
The horospheres in $\mathbb{H}^{3}$ with centres at $\gamma (\infty )$  can be defined as
the level set of the Busemann function $\varphi : \mathbb{H}^{3}\rightarrow \mathbb{R}$ associated with $\gamma$ and defined by
\begin{equation}
\varphi (x)\doteq \lim_{t\rightarrow \infty }d_{\mathbb{H}^{3}}(x,\gamma (t))-t,
\label{Busemann}
\end{equation}
where $d_{\mathbb{H}^{3}}(\,\,,\,\,)$ denotes the hyperbolic distance. Thus, to each points at infinity $\gamma (\infty )\in \partial \mathbb{H}^{3}$ is associated a foliation of $\mathbb{H}^{3}$ by horospheres which are the
level sets of the Busemann function. In particular, two horospheres with centre at the same
point at infinity are at a constant distance. Note that, as a set, the horospheres can be parametrized by future-pointing null vectors belonging to the future light-cone 
\begin{equation}
\mathbb{L}^{+}\bigskip \doteq \left\{ \overrightarrow{x}\doteq \left(
x^{0},x^{1},x^{2},x^{3}\right) \,\,|\;\left\langle \overrightarrow{x},%
\overrightarrow{x}\right\rangle =0,\;x^{0}>0\right\} 
\end{equation}
by identifying the generic horosphere $\Sigma _{w}$ with the intersection between $\mathbb{H}^{3}$ and the null hyperplane 
$\left\langle \overrightarrow{y},\overrightarrow{w}%
\right\rangle =-1$ defined by the null vector $\overrightarrow{w}$, {\emph{i.e.},}
\begin{equation}
\overrightarrow{w}\longmapsto \Sigma _{w}\doteq \left\{ y\in \mathbb{H}%
^{3}\bigskip \;|\;\left\langle \overrightarrow{y},\overrightarrow{w}%
\right\rangle =-1, \,\, \left\langle \overrightarrow{w},\overrightarrow{w}%
\right\rangle =0 \right\} .
\end{equation}
Such an identification allows to associate a natural functional with any pair of horospheres $\Sigma _{u}$ and $\Sigma _{v}$ according to 
\begin{equation}
\lambda \left( \Sigma _{u},\Sigma _{v}\right) \doteq \sqrt{-\left\langle 
\overrightarrow{u},\overrightarrow{v}\right\rangle }.
\label{lambdaLength}
\end{equation}
the quantity   $\lambda \left(
\Sigma _{u},\Sigma _{v}\right) $ defines the {\emph{lambda length}} \cite{Penner1} between $\Sigma _{u}$ and $\Sigma _{v}$.
If $\gamma (p,q)$ denotes the unique geodesic in $\mathbb{H}^{3}$ connecting
the respective centers $p\ $and $\ q$ of $\Sigma _{u}$ and $\Sigma _{v}$,
then $\lambda \left( \Sigma _{u},\Sigma _{v}\right) $\ \ can be related do
the signed geodesic distance $\delta (u,v)$ \ between the intersection points $%
\gamma (p,q)\cap \Sigma _{u}$ and $\gamma (p,q)\cap \Sigma _{v}$,  according to
\begin{equation}
\lambda \left( \Sigma _{u},\Sigma _{v}\right)=\sqrt{2\,e^{\delta (u,v)}} ,
\label{horodistance}
\end{equation}
($\delta
(u,v)$ is by convention $<0$ if $\Sigma _{u}$ and $\Sigma _{v}$ cross each
other).

\medskip

\noindent In order to discuss the connection between Regge triangulations and hyperbolic geometry,
it will be convenient to represent $\mathbb{H}^{3}$ by the upper half-space
model $\mathbb{H}^{3,+}_{up}$\ , \emph{i.e.} as the open upper half space $%
\left\{ (X,Y,Z)\in \mathbb{R}^{3}\;|\;Z>0\right\} $ \ endowed with the
Poincar\'{e} metric $Z^{-2}(dX^{2}+dY^{2}+dZ^{2})$. The boundary of $\mathbb{%
H}^{3}$ is here provided by $\partial \mathbb{H}^{3,+}_{up}=(\mathbb{R}%
^{2}\times \left\{ 0\right\} )\cup \left\{ \infty \right\} $, and, up to
isometries,  we can always map a given point $p$  to $\infty $. Geodesics 
in the half-space model are obtained by parametrization of vertical lines $%
\left\{ x\right\} \times \mathbb{R}_{+}$ and circles orthogonal to $\mathbb{R%
}^{2}\times \left\{ 0\right\} $. In particular, since geodesics with end
point $\infty $ are vertical lines, it easily follows that in $\mathbb{H}^{3,+}_{up}$
the horospheres (centered at $\infty $) are horizontal hyperplanes. 
It is also worthwhile recalling that the hyperbolic distance between two
points $p$, and $q$ $\in \mathbb{H}^{3}$ is explicitly provided in $\mathbb{H}^{3,+}_{up}$ by
\begin{equation}
d_{\mathbb{H}^{3}}(p,q)=2\tanh ^{-1}\left[ \frac{%
(X_{p}-X_{q})^{2}+(Y_{p}-Y_{q})^{2}+(Z_{p}-Z_{q})^{2}}{%
(X_{p}-X_{q})^{2}+(Y_{p}-Y_{q})^{2}+(Z_{p}+Z_{q})^{2}}\right] ^{\frac{1}{2}}.
\label{hyperdistance}
\end{equation}
In particular, if we take any two geodesics $l_{1}$ and $l_{2}$ with end-point $\infty $ and evaluate their hyperbolic distance $d_{\mathbb{H}^{3}}(l_{1},l_{2})$ along the horospheres $\Sigma _{1}\doteq \{z=t_{1}\}$ and $\Sigma _{2}\doteq \{z=t_{2}\}$, with $t_{2}>t_{1}$, separated by a distance $d_{\mathbb{H}^{3}}(\Sigma _{1},\Sigma _{2})$, then we get the useful relation
\begin{equation}
d_{\mathbb{H}^{3}}(l_{1},l_{2})|_{\Sigma _{2}}=d_{\mathbb{H}^{3}}(l_{1},l_{2})|_{\Sigma _{1}}\, e^{d_{\mathbb{H}^{3}}(\Sigma _{1},\Sigma _{2})}.
\label{reldistance}
\end{equation}

\bigskip

\noindent Let $\sigma _{hyp}^{3}\doteq (v^{0}(0),v^{0}(k),v^{0}(h),v^{0}(j))$ be an
ideal simplex in $\mathbb{H}^{3,+}_{up}$, \emph{i.e.}, a simplex whose faces are hyperbolic triangles, edges are geodesics, and with vertices lying on $\partial \mathbb{H}^{3,+}_{up}$. In order to describe the basic
properties of  $\sigma _{hyp}^{3}$ recall that, up to isometries of $\mathbb{H}^{3,+}_{up}$,
 we can always assume that  one of its four vertices, say $v^{0}(0)
$,  is at the point $\infty $ whereas the remaining three $\ v^{0}(k)$, $\ v^{0}(h)$, and $\ v^{0}(j)$ lie on the circumference intersection of $\mathbb{R}^{2}\times\{0\}$ with a Euclidean half-sphere $\mathbb{D}_{r}^{2}$ of radius $r$
and centre $c\in \left\{ (X,Y,Z)\in \mathbb{R}\,^{3}\,|\,Z=0\right\} $. \
Note that $\mathbb{D}_{r}^{2}$ inherits from $\mathbb{H}^{3,+}_{up}$ the structure of  a two-dimensional hyperbolic space and that, consequently the simplex $\sigma _{hyp}^{2}\doteq (v^{0}(k),v^{0}(h),v^{0}(j))$, providing the two-dimensional face of $\sigma _{hyp}^{3}\doteq (v^{0}(0),v^{0}(k),v^{0}(h),v^{0}(j))$ opposite to the vertex $v^{0}(0)\simeq \infty $, is itself an ideal simplex in $\mathbb{D}_{r}^{2}$. \
Denote by $\Delta_{\infty } (v^{0}(0))$\ the intersection between $\sigma _{hyp}^{3}$
and a horosphere $\Sigma _{\infty }$ centered at $v^{0}(0)\doteq \infty $
and sufficiently near to $v^{0}(0)$. Since all horospheres are congruent, $\Sigma _{\infty }$ can be mapped
onto a horizontal \ plane $z=t$ $\subset \mathbb{H}^{3,+}_{up}$ \ \ by a
conformal mapping fixing $\infty $, to the effect that $\Delta_{\infty } (v^{0}(0))$
is a Euclidean triangle $\ T_{\infty }(\sigma _{hyp}^{3})\equiv \sigma ^{2}(k,h,j)$ in the plane of the
horosphere. This latter remark implies that the vertex angles $(\theta _{jkh},\theta _{khj},\theta _{hjk})$ of 
$\ T_{\infty }(\sigma _{hyp}^{3})$ can be identified with the inner dihedral
angles  at the three edges $v^{1}(\infty ,k)$, $v^{1}(\infty ,h)$, and $v^{1}(\infty ,j)$ of $\ \sigma _{hyp}^{3}$%
, \emph{i.e.},
\begin{gather}
\theta _{jkh}\longmapsto \phi  _{\infty k}\doteq \angle \left[
v^{2}(0,j,k),v^{2}(0,k,h)\right] , \\
\theta _{khj}\longmapsto \phi _{\infty h}\doteq \angle \left[
v^{2}(0,k,h),v^{2}(0,h,j)\right] ,  \notag \\
\theta _{hjk}\longmapsto \phi _{\infty j}\doteq \angle \left[
v^{2}(0,h,j),v^{2}(0,j,k)\right] ,  \notag
\end{gather}
where $v^{2}(.,.,.)$\ denote the faces of $\sigma _{hyp}^{3}$.\ 
It is easy to prove, again by intersecting $\sigma _{hyp}^{3}$ with horospheres $\Sigma_{k}$, $\Sigma_{h}$, $\Sigma_{j}$ centered and sufficiently near to the respective vertices $v^{0}(k)$, $v^{0}(h)$, $v^{0}(j)$, that dihedral angles along opposite edges in $\sigma
_{hyp}^{3}$ are pairwise equal $\phi _{\infty k}=\phi _{hj}$,  $\phi
_{\infty h}=\phi _{jk}$, $\phi _{\infty j}=\phi _{kh}$. This implies that the (Euclidean) triangles cut by the horospheres $\Sigma_{k}$, $\Sigma_{h}$, $\Sigma_{j}$ are all similar to $\ T_{\infty }(\sigma _{hyp}^{3})$. In particular, note that the geometrical realizations of the simplices
\begin{gather}
\sigma _{hyp}^{2}(k,h,j)\doteq v^{2}(k,h,j),\\
\sigma _{hyp}^{2}(\infty, k,h)\doteq v^{2}(0,k,h),\notag \\
\sigma _{hyp}^{2}(\infty, j,k)\doteq v^{2}(0,j,k), \notag
\end{gather}
are  ideal triangles in $\mathbb{H}^{\,3}$. It follows that the above
construction is independent from the choice of which of the four vertices of $\sigma _{hyp}^{3}$ is mapped to $\infty $ and we can parametrize \ the
ideal tetrahedra $\sigma _{hyp}^{3}$ in $\mathbb{H}^{3,+}_{up}$ in terms of the
similarity class $[\sigma ^{2}(k,h,j)]$ of the associated \ Euclidean triangle $T(\sigma _{hyp}^{3})
$: any two ideal tetrahedra $\sigma _{hyp}^{3}$ in $\mathbb{H}^{3,+}_{up}$ are
congruent iff the associated triangles $T(\sigma _{hyp}^{3})$ are similar. This is in line with the basic property of  $\mathbb{H}^{\,3}$ according to which if a diffeomorphism of $\mathbb{H}^{\,3}$ preserves angles then it also preserves lengths.\

\subsection{\protect\bigskip Horospheres and twistors}

\noindent To conclude this brief r\'{e}sum\'{e} of hyperbolic geometry, let us observe that if we mark a point $P$ on the horosphere $\Sigma _{w}$, then there is a characterization of $(\Sigma _{w},P)$ in terms of {\emph{null twistors}} which will be relevant in what follows. To fix notation, \  let $\sigma _{0}^{AA^{\prime }}\doteq \delta^{AA^{\prime }}$ \ and denote by   
\begin{equation}
\sigma _{1}^{AA^{\prime }}\doteq \left( 
\begin{tabular}{ll}
$0$ & $1$ \\ 
$1$ & $0$%
\end{tabular}
\right) ,\,\,\sigma _{2}^{AA^{\prime }}\doteq \left( 
\begin{tabular}{ll}
$0$ & $i$ \\ 
$-i$ & $0$%
\end{tabular}
\right) ,\,\,\sigma _{3}^{AA^{\prime }}\doteq \left( 
\begin{tabular}{ll}
$1$ & $0$ \\ 
$0$ & $-1$%
\end{tabular}
\right) ,
\end{equation}
the Pauli matrices. Let \ $l_{w}$ the null line passing through the marked point $P\in \Sigma _{w}$ and contained
in the plane $\left\langle \overrightarrow{y},\overrightarrow{w}%
\right\rangle =-1$ whose intersection with $\mathbb{H}^{3}$ defines $\Sigma _{w}$.  \  Let us represent the coordinates $y^{k}$ of 
$P$ in terms of the Hermitian matrix 
\begin{equation}
P_{w}\longmapsto y^{AA^{\prime }}=\,\,\frac{1}{\sqrt{2}}\,y^{k}\,\sigma _{k}^{AA^{\prime }} =\frac{1}{\sqrt{2}}\left( 
\begin{tabular}{ll}
$y^{0}+y^{3}$ & $y^{1}+i\,y^{2}$ \\ 
$y^{1}-i\,y^{2}$ & $y^{0}-y^{3}$%
\end{tabular}
\right)
\end{equation}
and describe the null future-pointing vector $\overrightarrow{w}\in \mathbb{%
L}^{+}$ in terms of  a two components $SL(2,\mathbb{C})$ spinor $\xi \,^{A}$  as
$\overrightarrow{w}$ $\longleftrightarrow \xi \,^{A}\,%
\overline{\xi }\,^{A^{\prime }}$, (as usual, a representation which is unique up to a phase conjugation, {\emph{i.e.}}, $\xi \,^{A}\mapsto e^{i\,\varphi }\xi \,^{A}$). In terms of these quantities
we can associate with the pair $(\Sigma _{w}, P)$ the null
twistor 
\begin{equation}
(\Sigma _{w}, P)\longmapsto  W\,^{\Lambda }\doteq \left( \xi ^{A},\,\eta _{A^{\prime }}\right) \in \mathbb{C}P^{3}
\label{twistor}
\end{equation}
where $\Lambda  =1,2,3,4$ and  $\eta _{A^{\prime }}\doteq -i\,\xi
\,^{A}y_{AA^{\prime }}$ is the moment of $\xi \,^{A}$, \ (with respect to
the origin $O$), evaluated at the marked point  $P$. Conversely, from $\eta _{A^{\prime }}\doteq -i\,\xi
\,^{A}y_{AA^{\prime }}$  we have
\begin{equation}
y_{AA^{\prime }}=i\left(\xi\,^{B}\overline{\eta }\,_{B}\right)^{-1}\,\,\overline{\eta }\,_{A}\,\,{\eta }\,_{A^{\prime }},
\label{inversion}
\end{equation}
and we can parametrize the null line \ $l_{w}$  through $y_{AA^{\prime }}$ according to
\begin{equation}
t\longmapsto X_{AA^{\prime }}(t)=y_{AA^{\prime }}+t\,\,\overline{\xi  }\,_{A}\,\,{\xi }\,_{A^{\prime }}. 
\end{equation}
Note that, since there is a unique hyperbolic geodesics $\gamma $
passing through $P\in \Sigma _{w}$ with endpoint $\gamma (w)$, centre of $\Sigma _{w}$, the twistor $W^{A}$ can be equivalently thought of as representing $\gamma $ or, by duality, the Busemann function (\ref{Busemann}) associated with $\gamma $.
This correspondence between twistors and geodesics in $\mathbb{H}^{3}$ is particularly useful when dealing with two distinct horospheres $\Sigma _{w_{k}}$ and $\Sigma _{w_{h}}$, respectively represented by  $\xi \,^{A}(k)\,%
\overline{\xi }\,^{A^{\prime }}(k)$  and $\xi \,^{A}(h)\,%
\overline{\xi }\,^{A^{\prime }}(h)$. In such a case, to the ordered pair  $(\Sigma _{w_{k}},\Sigma _{w_{h}})$ we can associate the null twistor
\begin{equation}
(\Sigma _{w_{k}},\Sigma _{w_{h}})\longmapsto W\,^{\Lambda  }(k,h)\doteq \left( \xi ^{A}(k),\, \overline{\xi}
\,_{A^{\prime }}(h)     \right) \in \mathbb{C}P^{1}\times \mathbb{C}P^{1}\setminus \Delta
\label{twisthoro}
\end{equation}  
where $\Delta $ is the diagonal in $\mathbb{C}P^{1}\times \mathbb{C}P^{1}$, (which must be removed since the horospheres are distinct). Similarly, we have the correspondence  
\begin{equation}
(\Sigma _{w_{h}},\Sigma _{w_{k}})\longmapsto W\,^{\Lambda  }(h,k)\doteq \left( \xi ^{A}(h),\, \overline{\xi}
\,_{A^{\prime }}(k)     \right) 
\label{twisthoro2}
\end{equation} 
for the pair $(\Sigma _{w_{h}},\Sigma _{w_{k}})$ in reversed order.
The characterizations (\ref{twisthoro}) and (\ref{twisthoro2}) are twistorial in the sense that
 $W^{\Lambda  }(h,k)$ and \ $W^{\Lambda  }(k,h)$ are incident, \emph{%
i.e.},
\begin{equation}
W^{\Lambda  }(k,h)\,\overline{W}_{\Lambda  }(h,k)=\xi ^{A}(k)\xi \,_{A}(k)+\overline{%
\xi }_{A^{\prime }}(h)\overline{\xi }\,^{A^{\prime }}(h)=0
\end{equation}
where $W^{\Lambda  }(k,h)\,\overline{W}_{\Lambda  }(h,k)$ denotes the
(pseudo-hermitian) inner product in twistor space. This implies that
 we can find two  null lines $l_{k}$ and $l_{h}$, with respective tangent vectors
$\xi ^{A}(k)\overline{\xi }\,^{A^{\prime }}(k)$ and $\xi ^{A}(h)\overline{\xi }\,^{A^{\prime }}(h)$, intersecting each other at a point $x_{AA^{\prime }}(k,h)$  such that 
\begin{equation}
\overline{\xi}\,_{A^{\prime }}(h)=-i\,\xi
\,^{A}(k)x_{AA^{\prime }}(k,h),
\end{equation} 
\begin{equation}
\overline{\xi}\,_{A^{\prime }}(k)=-i\,\xi
\,^{A}(h)x_{AA^{\prime }}(k,h).
\end{equation}
Formally we can write
\begin{equation}
x_{JJ^{\prime }}(k,h)=\frac{i}{\xi ^{A}(k)\xi \,_{A}(h)}\left( \xi \,_{J}(k)%
\overline{\xi }_{J^{\prime }}(k)+\xi \,_{J}(h)\overline{\xi }_{J^{\prime
}}(h)\right).
\end{equation}
In order to characterize the point $x_{AA^{\prime }}(k,h)$ in geometrical terms, observe that 
given any two distinct horospheres $\Sigma _{w_{k}}$ and $\Sigma _{w_{h}}$ there is a unique parametrized hyperbolic geodesic $t\mapsto \gamma(k,h)(t)$ connecting them. Such a geodesic  marks a unique point $y_{(\Sigma_{ k},h)}$ $\doteq $ $\gamma(k,h) \cap \Sigma _{w_{k}}$ on $\Sigma _{w_{k}}$, (similarly there is a unique point $y_{(\Sigma _{h},k)}$ $\doteq $ $\gamma(h,k) \cap \Sigma _{w_{h}}$ intercepted on $\Sigma _{w_{h}}$ by the orientation-reversed geodesic $\gamma(h,k)$). We let  $l_{k}$ be the null line passing through $y_{(\Sigma_{ k},h)}$ with tangent vector
$\xi ^{A}(k)\overline{\xi }\,^{A^{\prime }}(k)$, and $l_{h}$  the null line passing through $y_{(\Sigma_{ h},k)}$ with tangent     $\xi ^{A}(h)\overline{\xi }\,^{A^{\prime }}(h)$. Both such lines lie in a two-dimensional hyperplane  $\subset M^{4}$ passing through the origin and whose intersection with $\mathbb{H}^{3}$ traces the geodesic $\gamma(k,h)$. Since the horospheres $\Sigma _{w_{k}}$ and $\Sigma _{w_{h}}$ are distinct, the lines $l_{k}$ and $l_{h}$ necessarily intersect in a point which provides the required 
$x_{AA^{\prime }}(k,h)$. 
Recall that in terms of the spinorial representation $\overrightarrow{w}$ $%
\longleftrightarrow \xi \,^{A}\,\overline{\xi }\,^{A^{\prime }}$ and $%
\overrightarrow{v}$ $\longleftrightarrow \zeta \,^{A}\,\overline{\zeta }%
\,^{A^{\prime }}$ of  two future-pointing null vector $\overrightarrow{w}$
and \ $\overrightarrow{v}$  we can write
\begin{equation}
-\left\langle \overrightarrow{v},\overrightarrow{w}\right\rangle =\frac{1}{2}%
\zeta \,^{A}\,\overline{\zeta }\,^{A^{\prime }}\xi \,^{B}\,\overline{\xi }%
\,^{B^{\prime }}\epsilon _{AB}\,\epsilon _{A^{\prime }B^{\prime }}=\frac{1}{2%
}\zeta \,_{B}\,\overline{\zeta }\,_{B^{\prime }}\,\xi \,^{B}\,\overline{\xi }%
\,^{B^{\prime }},
\end{equation}
where $\epsilon _{AB}$ is the antisymmetric (symplectic) 2-form on spinor
space (chosen so that $\epsilon _{01}=1$ in the selected spin frame), and where spinorial indices are
lowered and raised \ via $\zeta \,_{B}=\zeta \,^{A}\epsilon _{AB}$ and $%
\zeta \,^{A}=\epsilon ^{AB}\zeta \,_{B}$. \ Applying this to the null
vectors $\overrightarrow{w}(k)$ and $\overrightarrow{w}(h)$, defining the two
horospheres $\Sigma _{w_{k}}$ and $\Sigma _{w_{h}}$, we get the twistorial
expression of the corresponding $\lambda $-length  
\begin{equation}
\lambda (\Sigma _{w_{k}},\Sigma _{w_{h}})=\sqrt{\frac{1}{2}\,\,\xi _{B}(k)\,%
\overline{\xi }_{B^{\prime }}(k)\,\xi\, ^{B}(h)\,\overline{\xi }\,^{B^{\prime
}}(h)}\doteq \frac{1}{\sqrt{2}}\left|\left|\xi _{B}(k)\xi\, ^{B}(h) \right|\right|.
\label{spinnorm}
\end{equation}
In terms of the geodesic $\gamma(k,h)$ we are basically describing the well-known twistor correspondence \cite{Baird} between the geodesic field of $\mathbb{H}^{3}$ and the (mini)twistor space $\mathbb{C}P^{1}\times \mathbb{C}P^{1}\setminus \Delta $.

\subsection{\protect\bigskip The computation of Lambda-lengths}
 \label{LLengths}

\noindent A key step in discussing the relation among Regge triangulations, twistor theory, and hyperbolic geometry involves the computation of the lambda-lengths (\ref{lambdaLength}) in terms of the Euclidean lengths of the edges of $\sigma ^{2}(k,h,j)$. To this end, we consider horospheres  $\Sigma _{k}$, $\Sigma _{h}$,$\Sigma _{j}$ sufficiently near to the vertices $v^{0}(k)$, $v^{0}(h)$, $v^{0}(j)$ of $\sigma_{hyp} ^{2}(k,h,j)$. We start by evaluating the lambda-lengths (\ref{lambdaLength}) along the vertical geodesics connecting $v^{0}(0)\simeq \infty$ with the triangle $\sigma _{hyp}^{2}(k,h,j)$. By applying 
(\ref{hyperdistance}) and (\ref{horodistance}) we get
\begin{eqnarray}
\lambda \left( \Sigma _{\infty },\Sigma _{k}\right)  &=&\sqrt{2\frac{t%
}{z_{k}}}, \\
\lambda \left( \Sigma _{\infty },\Sigma _{h}\right)  &=&\sqrt{2\frac{t%
}{z_{h}}},  \notag \\
\lambda \left( \Sigma _{\infty },\Sigma _{j}\right)  &=&\sqrt{2\frac{t%
}{z_{j}}},  \notag
\end{eqnarray}
where $z=z_{k}$, $z=z_{h}$, and $z=z_{j}$ respectively define the $z$ coordinates of the
intersection points between the horospheres $\Sigma _{k}$, $\Sigma
_{h}$, $\Sigma _{j}$ and the corresponding vertical geodesics.
Consider now the intersection of the ideal triangle $\sigma _{hyp}^{2}(\infty
,k,h)$ with each of the horospheres $\Sigma
_{\infty }$, $\Sigma _{k}$, and $\Sigma _{h}$. Each such an intersection characterizes a
corresponding horocyclic segment  $\digamma _{\infty }$, $\digamma
_{k}$, $\digamma _{h}$  whose hyperbolic length defines
(twice) the $h$-lenght of the horocyclic segment. In particular, the
horocyclic segment traced by $\sigma _{hyp}^{2}(\infty
,k,h)\cap \Sigma _{\infty }$ is the side $\sigma ^{1}(k,h)$ of
the Euclidean triangle $\sigma ^{2}(k,h,j)$. According to (\ref{hyperdistance}), its $h$-lenght
is provided by 
\begin{equation}
K_{t}\,(k,h)=\tanh ^{-1}\sqrt{\frac{l^{2}(k,h)}{l^{2}(k,h)+4t^{2}}}.
\end{equation}
On the other hand, the \ horocyclic segment $\sigma ^{1}(k,h)$ is opposite
to the geodesic segment intercepted by the horospheres $\Sigma _{k}$,
and $\Sigma _{h}$ along the hyperbolic edge $\ \sigma
_{hyp}^{1}(k,h)$. The lambda-length of this segment is $%
\lambda \left( \Sigma _{k},\Sigma _{h}\right) $, and according
to a result by R. Penner \cite{Penner1}, (Proposition 2.8), among these quantities there holds the relation  
\begin{equation}
K_{t}\,(k,h)=\frac{\lambda \left( \Sigma _{k},\Sigma _{h}\right) }{%
\lambda \left( \Sigma _{\infty },\Sigma _{k}\right) \lambda \left(
\Sigma _{\infty },\Sigma _{h}\right) },
\label{Hlength}
\end{equation}
from which we get 
\begin{equation}
\lambda \left( \Sigma _{k},\Sigma _{h}\right) =\frac{2t}{\sqrt{%
z_{k}z_{h}}}\,\tanh ^{-1}\sqrt{\frac{l^{2}(k,h)}{l^{2}(k,h)+4t^{2}}.}
\end{equation}
Similarly, we compute 
\begin{equation}
\lambda \left( \Sigma _{h},\Sigma _{j}\right) =\frac{2t}{\sqrt{%
z_{h}z_{j}}}\,\tanh ^{-1}\sqrt{\frac{l^{2}(h,j)}{l^{2}(h,j)+4t^{2}},}
\end{equation}
\begin{equation}
\lambda \left( \Sigma _{j},\Sigma _{k}\right) =\frac{2t}{\sqrt{%
z_{j}z_{k}}}\,\tanh ^{-1}\sqrt{\frac{l^{2}(j,k)}{l^{2}(j,k)+4t^{2}}.}
\end{equation}
Note that these relations must hold for any $t$, and if we take the limit as $t\longrightarrow \infty $ we easily find
\begin{equation}
\lambda \left( \Sigma _{k},\Sigma _{h}\right) =\frac{l(k,h)}{\sqrt{%
z_{k}z_{h}}},
\label{luno}
\end{equation}
\begin{equation}
\lambda \left( \Sigma _{h},\Sigma _{j}\right) =\frac{l(h,j)}{\sqrt{%
z_{h}z_{j}}},
\label{ldue}
\end{equation}
\begin{equation}
\lambda \left( \Sigma _{j},\Sigma _{k}\right) =\frac{l(j,k)}{\sqrt{%
z_{j}z_{k}}}.
\label{ltre}
\end{equation}
We can also compute the h-lengths associated with the decorated ideal triangle
$\ \sigma_{hyp}^{2}(k,h,j)$ and defined by
\begin{equation}
H\,(\Sigma _{k},\Sigma _{h})\doteq \frac{\lambda \left( \Sigma
_{k},\Sigma _{h}\right) }{\lambda \left( \Sigma
_{h},\Sigma _{j}\right) \lambda \left( \Sigma
_{j},\Sigma _{k}\right) },
\label{cinquantatre}
\end{equation}
\begin{equation}
H\,(\Sigma _{h},\Sigma _{j})\doteq \frac{\lambda \left( \Sigma
_{h},\Sigma _{j}\right) }{\lambda \left( \Sigma
_{j},\Sigma _{k}\right) \lambda \left( \Sigma
_{k},\Sigma _{h}\right) },
\end{equation}
\begin{equation}
H\,(\Sigma _{j},\Sigma _{k})\doteq \frac{\lambda \left( \Sigma
_{j},\Sigma _{k}\right) }{\lambda \left( \Sigma
_{k},\Sigma _{h}\right) \lambda \left( \Sigma
_{h},\Sigma _{j}\right) }.
\end{equation}
From (\ref{luno}) $\div $ (\ref{ltre}) we get 
\begin{equation}
H\,(\Sigma _{k},\Sigma _{h})=\frac{l(k,h)}{l(h,j)l(j,k)}\,z_{j},
\label{antasei}
\end{equation}
\begin{equation}
H\,(\Sigma _{h},\Sigma _{j})=\frac{l(h,j)}{l(j,k)l(k,h)}\,z_{k},
\end{equation}
\begin{equation}
H\,(\Sigma _{j},\Sigma _{k})=\frac{l(j,k)}{l(k,h)l(h,j)}\,z_{h}.
\end{equation}

\bigskip

\noindent Since the $\lambda $-length $\lambda (\Sigma _{k},\Sigma _{h})$ can also be
expressed in terms of the spinorial quantity $2^{-\frac{1}{2}}||\xi _{B}(k)\xi
^{B}(h)||$, (see(\ref{spinnorm})), we can use the above connection with the Euclidean lengths $%
\left\{ l(k,h)\right\} $ in order to provide, in terms of the spinorial norms $\left\{ ||\xi _{B}(k)\xi ^{B}(h)||\right\}$ and of the $\left\{l(k,h)\right\}$,
the $z$-coordinates $\left\{ z_{k}\right\} $ of the marked points on the
horospheres. \ Explicitly, from the expressions (\ref{cinquantatre}), (\ref{antasei}) of the h-length $%
H(\Sigma _{k},\Sigma _{h})$ and (\ref{spinnorm}) we get
\begin{equation}
H(\Sigma _{k},\Sigma _{h})=\sqrt{2}\,\,\frac{||\xi _{B}(k)\xi ^{B}(h)||}{||\xi
_{B}(h)\xi ^{B}(j)||\,\,||\xi _{B}(j)\xi ^{B}(k)||}=\frac{l(k,h)}{l(h,j)l(j,k)}%
\,\,z_{j},
\end{equation}
from which we compute
\begin{equation}
z_{j}=\sqrt{2}\,\,\frac{||\xi _{B}(k)\xi ^{B}(h)||}{||\xi _{B}(h)\xi ^{B}(j)||\,\,||\xi
_{B}(j)\xi ^{B}(k)||}\frac{l(h,j)l(j,k)}{l(k,h)},
\end{equation}
(by a cyclical permutation of $(k,h,j)$ we easily get
the expressions for $z_{h}$, and $z_{k}$). Note that the knowledge of the $%
\left\{ z_{k}\right\} $ provides, up to translations in $\mathbb{R}%
^{2}\times \left\{ 0\right\} $, the horospheres $\left\{ \Sigma _{k}\right\} 
$ in $\mathbb{H}_{up}^{3,+}$ which decorate the vertices of \ the ideal
triangle $\sigma _{hyp}^{2}(k,h,j)$, (the actual position of these
horospheres is defined by the corresponding null vector $\xi ^{B}(k)\overline{\,\xi }%
\,^{B^{\prime }}(k)\,$).  It follows from these remarks that from the
twistorial decoration of the Euclidean triangle  $\sigma ^{2}(k,h,j)$ we can
fully recover the horospherically decorated hyperbolic triangle $\sigma
_{hyp}^{2}(k,h,j)$. In other words we have the correspondance
\begin{gather}
\fbox{Euclidean triangles decorated with 
null pairwise incident twistors}\notag \\
\Updownarrow \notag \\ 
\fbox{Ideal hyperbolic triangles
decorated with horospheres} \notag
\end{gather}
This directly bring us to  discuss what kind of structure is induced in $%
\mathbb{H}_{up}^{3,+}$ by a twistorial field defined over the whole triangulation $\left| T_{l}\right|
\rightarrow M$.

\section{\protect\bigskip Regge triangulations in twistor space}
\label{RTITS}
\noindent  The geometrical analysis  of the previous paragraphs  implies that
to each  of the $N_{2}(T)$ Euclidean triangles $\sigma ^{2}(k,h,j)$ of $%
|T_{l}|\rightarrow M$ we can associate a ideal  tetrahedron $\sigma
_{hyp}^{3}(\infty ,k,h,j)$ in  $\mathbb{H}_{up}^{3,+}$ and an ideal triangle 
$\sigma _{hyp}^{2}(k,h,j)$ decorated with the
horocyclic sectors induced by a choice of horospheres $\Sigma _{k}$, $%
\Sigma _{h}$, $\Sigma _{j}$.
Note that the decoration of the vertices $v^{0}(k)$, $v^{0}(h)$, and $%
v^{0}(j)$ actually exploits the data of $(z_{k},\,\Sigma _{k})$, $(z_{h},\,\Sigma _{h})$, and $%
(z_{j},\,\Sigma _{j})$ where the points $z_{k}$, $z_{h}$, and $z_{j}$ belongs to the respective horospheres and determine the geodesic $\gamma $ in $\mathbb{H}_{up}^{3,+}$ whose endpoint $\gamma (\infty )$ is the centre of  $\Sigma _{\infty }$.
According to (\ref{twistor}) this decoration of $\sigma _{hyp}^{2}(k,h,j)$  can be thought of as induced by the twistorial decoration of the vertices  of the Euclidean triangle $\sigma ^{2}(k,h,j)$ defined by the map
\begin{gather}
\left\{ \sigma ^{0}(i)\right\} _{V(T)}\longrightarrow \,\mathbb{C}P^{3} \label{tcorrespondence} \\
\sigma ^{0}(k)\longmapsto \left( \xi ^{A}(k),\eta _{A^{\prime }}(k)\right), \notag 
\end{gather}
which associates with each vertex $\sigma ^{0}(k)\in $  $|T_{l}|\rightarrow M$ \
the null twistor describing the marked horosphere $(\Sigma _{k},z_{k})$. Equivalently, we can use the decoration defined by the null twistor $W^{\Lambda  }(k,\infty )$ describing the geodesic $\gamma $, {\emph{i.e.}}, 
\begin{gather}
\left\{ \sigma ^{0}(i)\right\} _{V(T)}\longrightarrow \,\mathbb{C}%
P^{1}\times \mathbb{C}P^{1}\backslash \Delta \subset \,\mathbb{C}P^{3}  \label{tcorrespondencebis} \\
\sigma ^{0}(k)\longmapsto W^{\Lambda  }(k,\infty )\doteq \left( \xi ^{A}(k),\overline{\xi} _{A^{\prime }}(\infty )\right) ,\notag 
\end{gather}
where ${\xi} ^{A}(\infty )\overline{\xi} ^{A^{\prime }}(\infty )$ is the null vector defining the horosphere $\Sigma _{\infty }$.
Whichever representative we chose, the edges of $\sigma ^{2}(k,h,j)$ carry an induced twistorial decoration defined by
\begin{gather}
\left\{ \sigma ^{1}(k,h)\right\} _{E(T)}\longrightarrow \mathbb{C}%
P^{1}\times \mathbb{C}P^{1}\backslash \Delta \subset \,\mathbb{C}P^{3} 
\label{edcorrespondence} \\
\sigma ^{1}(k,h)\longmapsto W^{\Lambda  }(k,h)\doteq \left( \xi ^{A}(k),\overline{\xi} _{A^{\prime }}(h)\right) ,
\notag
\end{gather}
which to each oriented edge $\sigma
^{1}(k,h)$ of $|T_{l}|\rightarrow M$
associates the null twistor  $( \xi ^{A}(k),\overline{\xi} _{A^{\prime }}(h)) $
describing the parametrized  geodesic $\gamma (k,h)$. It is worthwhile noticing that
the massless twistor fields defined on $|T_{l}|\rightarrow M$  by (\ref{tcorrespondencebis}) and (\ref{edcorrespondence}) can be equivalently thought of as providing a geometrical realization of an immersion of the random Regge triangulation 
$|T_{l}|\rightarrow M$ in the quadric $\mathbb{C}%
P^{1}\times \mathbb{C}P^{1}\backslash \Delta $ in twistor space $\mathbb{C}P^{3}$. 

\bigskip

\noindent As we have seen in  paragraph \ref{LLengths}, the above twistorial decoration allows to associate with each Euclidean triangle $\sigma ^{2}(k,h,j)$ a corresponding ideal triangle $\sigma _{hyp}^{2}(k,h,j)$ with a horocyclical decoration of the vertices, which can be recovered in terms of the Euclidean lenghts $\{l(k,h)\}$ of $\sigma ^{2}(k,h,j)$ and of the spinorial norms $\left\{ ||\xi _{B}(k)\xi ^{B}(h)||\right\} $.
Clearly, there is more in such a correspondence, and in particular one is naturally led to 
explore the possibility of glueing the ideal triangles $\{\sigma _{hyp}^{2}(k,h,j)\}$ in the same combinatorial pattern defined by $|T_{l}|\rightarrow M$. This must be done in such a way that the twistor fields on $|T_{l}|\rightarrow M$ provide a consistent horocyclical decoration of the vertices of the ideal triangulation defined by $\{\sigma _{hyp}^{2}(k,h,j)\}$. In performing such an operation one must take care of three basic facts:\, {\emph{(i)}}\,\, ideal triangles are rigid since any two of them are congruent; {\emph{(ii)}}\,\,  the adjacent sides of two ideal triangles can be identified up to the freedom of performing an arbitrary traslation along the edges, (each edge of an ideal triangle is isometric to the real line, its hyperbolic lenght being infinite, and two adjacent edges may freely slide one past another); {\emph{(iii)}}\,\, Since $\mathbb{H}^{3}$ is a space of left cosets of $SU(2)$ in $Sl(2,\mathbb{C})$, the identification of the marked point $z_{k}$ on the horosphere $(\Sigma _{k},z_{k})$, associated with a vertex $\sigma ^{0}(k)$, is only  defined up to the action of $SU(2)$.
These translational and $SU(2)$ degrees of freedom can be exploited in order to specifing how the decoration provided  by the horocyclic sectors in an ideal triangle is extended to the adjacent ideal triangle. Within such a set-up, let us
consider the star $Star[\sigma ^{0}(k)]$ of a generic vertex $\sigma
^{0}(k)$ over which $q(k)$ triangles $\sigma ^{2}(k,h_{\alpha },h_{\alpha +1})$ are incident. Let us label the corresponding set of hyperbolic ideal triangles by $\sigma ^{2}_{hyp}(k,h_{\alpha },h_{\alpha +1})$, $\alpha =1,...,q(k)$, with\ \ $h_{\alpha }=h_{\beta }$ if \ $\beta =\alpha \;\func{mod}\,q(k)$.
The natural hyperbolic structure on 
\begin{equation}
P^{2}(k)\doteq \bigcup_{\alpha =1}^{q(k)}\sigma _{hyp}^{2}(k,h_{\alpha
},h_{\alpha +1})-\left\{ v^{0}(k)\right\}, 
\end{equation}
($v^{0}(k)$ being the vertex associated with $\sigma ^{0}(k)$),
induces a similarity structure on the link associated with $v^{0}(k)$%
\begin{equation}
link\,\,\left[ v^{0}(k)\right] \doteq \bigcup_{\alpha =1}^{q(k)}\sigma
_{hyp}^{1}(h_{\alpha },h_{\alpha +1}),
\end{equation}
which characterizes, as $k$ varies, the hyperbolic surface one gets by glueing the hyperbolic triangles $\sigma _{hyp}^{2}(k,h_{\alpha
},h_{\alpha +1})$. To determine such a similarity structure, let us 
consider a triangle, say $\sigma
_{hyp}^{2}(k,h_{1},h_{2})$, in $P^{2}(k)$, and  let  
$\digamma _{k}^{h_{1}}$ be the oriented horocyclic segment cut in $\sigma
_{hyp}^{2}(k,h_{1},h_{2})$ by the horosphere $%
\Sigma _{k}$. This horocyclic segment can be extended, in a counterclockwise order, to the other $%
q(k)-1$ ideal triangles in the set $\{\sigma ^{2}_{hyp}(k,h_{\alpha },h_{\alpha +1})\}$ by requiring that such an extension meets orthogonally  each adjacent geodesic side of the $q(k)$ triangles considered. Since the horospheres are congruent and the identification between adjacent sides of ideal triangles is only defined up to a shift,  such an extension procedure generates a sequence of $q(k)$ horocyclic segments $\{\digamma _{k}^{h_{\alpha }}\}$ which
eventually re-enters the triangle $\sigma
_{hyp}^{2}(k,h_{1},h_{2})$ with a  horocyclic segment   $%
\widehat{\digamma }_{k}^{h_{q(k)}}$ which will be parallel
to $\digamma _{k}^{h_{1}}$ but not necessarily coincident with it.
The similarity structure is completely characterized by the Euclidean
similarity $f:\mathbb{R}\rightarrow \mathbb{R}$ which maps, along $\sigma
_{hyp}^{1}(h_{1},h_{2})$, the point $\digamma _{k}^{h_{1}}\cap \sigma
_{hyp}^{1}(h_{1},h_{2})$ to the point $\widehat{\digamma }%
_{k}^{h_{q(k)}}\cap \sigma _{hyp}^{1}(h_{1},h_{2})$. The horocycle curve $%
t\mapsto \digamma _{k}(t)$, $0\leq t\leq 2\pi $ closes up, \emph{i.e.}, $%
\digamma _{k}^{h_{1}}\cap \sigma _{hyp}^{1}(h_{1},h_{2})=$\ $\widehat{%
\digamma }_{k}^{h_{q(k)}}\cap \sigma _{hyp}^{1}(h_{1},h_{2})$ iff the
Euclidean length $|\digamma _{k}(t)|_{Euc}$ of \ $t\mapsto \digamma _{k}(t)
$ is $2\pi $, (note that $|\digamma _{k}(t)|_{Euc}$ is always a constant). In our case,\ \ the Euclidean length of the
horocycle curve $t\mapsto \digamma _{k}(t)$, $0\leq t\leq 2\pi $ is  given
by the conical defect $\Theta (k)=\Sigma _{\alpha =1}^{q(k)}\theta _{\alpha
+1,k,\alpha }$ supported at the vertex $\sigma ^{0}(k)\in |T_{l}|\rightarrow
M$.  Thus, the similarity ratio is given by $(\frac{\Theta (k)}{2\pi })$. Given such a ratio
one can compute, by exploiting (\ref{reldistance}), the signed hyperbolic distance between the horocycle
segments $\digamma _{k}^{h_{1}}$ and  $\widehat{\digamma }_{k}^{h_{q(k)}}$
according to
\begin{equation}
\mp  d_{\mathbb{H}^{3}}(\digamma _{k}^{h_{1}},\,\widehat{\digamma }_{k}^{h_{q(k)}})=\ln\frac{\Theta (k)}{2\pi }\doteq d[v^{0}(k)],
\label{reenter}
\end{equation}
where \ the sign is chosen to be positive
iff \ $\Theta (k)<2\pi $, {\emph{i.e.}} if   the\ horodisk sector bounded by $\digamma _{k}^{h_{1}}$ contains
the sector bounded by  $\widehat{\digamma }_{k}^{h_{q(k)}}$.
Note that the
number  $d[v^{0}(k)]$ does not depend from the initial choice of $\digamma_{v^{0}(k)}^{h_{1}}$, and is an invariant only related to the conical
defect \ $\Theta (k)$ supported at  the vertex $\sigma ^{0}(k)$ of $|T_{l}|\rightarrow M$. It can be identified with the invariant introduced by W. Thurston \cite{ThurstonB} in order to characterize the completeness of the hyperbolic structure of a surface obtained by gluing hyperbolic ideal triangles (the structure being complete iff the invariants $d[v^{0}(k)]$ are all zero for each ideal vertex $v^{0}(k)$).  A classical result by W. Thurston, (\cite{ThurstonB}, prop. 3.10.2), implies that the glueing of the $N_{2}(T)$ ideal triangles according to the procedure just described gives rise to an open hyperbolic surface $\Omega $ with geodesic boundaries. Each boundary component $\partial \Omega  _{k}$ is associated with a corresponding vertex $\sigma ^{0}(k)$ of $|T_{l}|\rightarrow M$, and has a length provided by 
\begin{equation}
\left| \partial \Omega  _{k}\right| =\left| d(v^{0}(k))\right| =\left| \ln \,%
\frac{\Theta (k)}{2\pi }\right| .
\end{equation}
These geodesic boundaries come also endowed with a $SU(2)$ holonomy which is generated by the twistors 
$\left( \xi ^{A}(k),\eta _{A^{\prime }}(k)\right)$ decorating each vertex $\sigma ^{0}(k)$. Explicitly, let
\begin{equation}
\left( \xi ^{A}(k),\,\eta _{A^{\prime }}(k)\right) _{\sigma ^{2}(k,h_{\alpha
},h_{\alpha +1})}
\end{equation}
denote the twistor associated with \ the marked horosphere \ $(\Sigma
_{k},z_{k})$ which decorates the vertex \ $v^{0}(k)$ of the triangle $%
\sigma_{hyp} ^{2}(k,h_{\alpha },h_{\alpha +1})$. If we denote by $I\doteq \frac{1}{%
\sqrt{2}}\delta _{AA^{\prime }}$\ the hermitian matrix corresponding to $%
(1,0,0,0)\in \mathbb{H}^{3}$, then we can set\ \ $\ \eta _{A^{\prime
}}(k)\doteq -i\,\xi \,^{A}(k)\,z_{AA^{\prime }}(k)$ where \ $z_{AA^{\prime
}}(k)$ is the $SL(2,\mathbb{C})$ matrix associated with the marked point $%
z_{k}$. When we move\ from the triangle $\sigma ^{2}(k,h_{\alpha },h_{\alpha
+1})$ to the adjacent one $\sigma ^{2}(k,h_{\alpha +1},h_{\alpha +2})$ the
corresponding group elements \ \ $z_{AA^{\prime }}(k)|_{\sigma ^{2}(k,h_{\alpha },h_{\alpha
+1})}$ and $z_{AA^{\prime }}(k)|_{\sigma ^{2}(k,h_{\alpha +1},h_{\alpha +2})%
\text{ \ \ }}$, being associated with the same coset $\{z_{k}\}\in $ $%
SL(2,\mathbb{C})\diagup SU(2)$, are related by 
\begin{equation}
z_{AA^{\prime }}(k)|_{\sigma ^{2}(k,h_{\alpha +1},h_{\alpha +2})\text{ \ \ }%
}=z_{AA^{\prime }}(k)|_{\sigma ^{2}(k,h_{\alpha },h_{\alpha
+1})}\,s(k,h_{\alpha +1}),
\end{equation}
where $s(k,h_{\alpha +1})\in SU(2)$, and where the labelling $(k,h_{\alpha
+1})$ refers to the edge $\sigma ^{1}(k,h_{\alpha +1})$ shared between the
two adjacent triangle $\sigma ^{2}(k,h_{\alpha },h_{\alpha +1})$ and $\sigma
^{2}(k,h_{\alpha +1},h_{\alpha +2})$. \ Since the locally Euclidean structure in each $Star[\sigma ^{0}(k)]$ is characterized by a $U(1)$ holonomy
$e^{i\,\Theta (k)}$, we require that $s(k,h_{\alpha +1})$ lies in the maximal torus in $SU(2)$, $\emph{%
i.e.}$  
\begin{equation}
s(k,h_{\alpha +1})=\,e^{i\sigma _{3}\,\theta _{\alpha +1,k,\alpha }}\doteq \left( 
\begin{tabular}{ll}
$e^{i\,\,\theta _{\alpha +1,k,\alpha }}$ & $0$ \\ 
$0$ & $e^{-i\,\,\theta _{\alpha +1,k,\alpha }}$%
\end{tabular}
\right) .
\end{equation}
Thus, by circling around the star \ $%
Star[\sigma ^{0}(k)]$\bigskip\ we get 
\begin{equation}
z_{AA^{\prime }}(k)|_{\sigma ^{2}(k,h_{q(k)},h_{1})\text{ \ \ }%
}=z_{AA^{\prime }}(k)|_{\sigma ^{2}(k,h_{1},h_{2})}\prod_{\alpha
=1}^{q(k)}s(k,h_{\alpha +1}),
\end{equation}
with $h_{q(k)+1}=h_{1}$. The group element defined by 
\begin{equation}
U_{k}\doteq \prod_{\alpha =1}^{q(k)}s(k,h_{\alpha +1})=e^{i\sigma _{3}\,\Theta (k)}
\end{equation}
provides the $SU(2)$ holonomy associated with the geodesic boundary $%
\partial \Omega  _{k}$. Associated with such a holonomy we have $
\mathfrak{su}(2)$-valued flat gauge potentials $A_{(k)}$ locally defined by 
\begin{equation}
A_{(k)}\doteq \frac{i}{4\pi }|\partial \Omega  _{k}|\gamma _{k}\left( \Theta (k)
\mathbf{\ \sigma }_{3}\right) \gamma _{k}^{-1}\left( \frac{d\zeta (k)}{\zeta
(k)}-\frac{d\overline{\zeta }(k)}{\overline{\zeta }(k)}\right) , 
\label{gpoten}
\end{equation}
where $\zeta (k)$ is a complex coordinate in a neighborhood of the boundary component $%
\partial \Omega  _{k}$, (defined by glueing to $\partial \Omega  _{k}$ a punctured disk $\in \mathbb{C}$ with coordinate $\zeta (k)$), and where $\gamma _{i}\in SU(2)$. 
We can sum up these remarks in the following
 
\begin{proposition}
A closed random Regge surface 
$(|T_{l}|\rightarrow M)$ whose vertex set $\{\sigma ^{0}(k)\}_{1}^{V(T)}$ is decorated with
the null twistor field $\sigma ^{0}(k)$ $\longmapsto$ $W^{\Lambda }(k,\infty )$ 
has a dual description as the
ideal triangulation $H\left( |T_{l}|\rightarrow M\right) $ of an open
hyperbolic surface $\Omega  \sim M/V(T)$ with geodesic boundaries $\partial \Omega  _{k}$ \ \ 
of length $| \partial \Omega  _{k}| =| \ln \,%
\frac{\Theta (k)}{2\pi }|$. To any such a boundary it is associated  a $SU(2)$ holonomy $U_{k}=e^{i\sigma _{3}\Theta (k)}$ generated by
a $\mathfrak{su}(2)$-valued flat gauge potential $A_{(k)}$.  
\end{proposition}

\subsection{\protect\bigskip Moduli kinematics of closed/open duality}

Let \ $\mathcal{T}_{g,N_{0}}(L)$ denote the Teichm\"{u}ller space of
hyperbolic surfaces $\Omega $ with geodesic boundary components of length
\begin{equation}
L=\left( L_{1},...,L_{N_{0}}\right) \doteq \left( |\partial \Omega
_{1}|,...,|\partial \Omega _{N_{0}}|\right) \in \mathbb{R}_{+}^{N_{0}}.
\end{equation}
Note that, by convention, a boundary component such that $|\partial \Omega
_{j}|=0$ is a cusp and moreover \ $\mathcal{T}_{g,N_{0}}(L=0)=\mathcal{T}%
_{g,N_{0}}$, where \ \ $\mathcal{T}_{g,N_{0}}$ is \ the Teichm\"{u}ller
space of hyperbolic surfaces with $N_{0}$ punctures, (with $6g-6+2N_{0}\geq 0
$). The elements of $\mathcal{T}_{g,N_{0}}(L)$ are marked Riemann surface
modelled on a surface $S_{g,N_{0}}$\ of genus $g$  \ with complete
finite-area metric  of constant Gauss curvature $-1$,\emph{\ }(and with $%
N_{0}$ geodesic boundary components $\partial S=\sqcup \partial S_{j}$ of
fixed length),\emph{\ \ i.e.}, a triple $(S_{g,N_{0}},\,f,\,\Omega )$ where $%
f:S_{g,N_{0}}\rightarrow \Omega $ is a quasiconformal homeomorphism, (the
marking map), which extends uniquely to a homeomorphism from $%
S_{g,N_{0}}\cup \partial S$ onto $\Omega \cup \partial \Omega $. Any two
such a triple $(S_{g,N_{0}},\,f_{1},\,\Omega _{(1)})$ and $%
(S_{g,N_{0}},\,f_{2},\,\Omega _{(2)})$ are considered equivalent iff there
is a biholomorphism $h:\Omega _{(1)}\rightarrow \Omega _{(2)}$ such that $%
f_{2}^{-1}\circ h\circ f_{1}$ $:$ \ $S_{g,N_{0}}\cup \partial S\rightarrow
S_{g,N_{0}}\cup \partial S$ is homotopic to the identity via continuous
mappings pointwise fixing $\partial S$.  For each given string  $L=\left(
L_{1},...,L_{N_{0}}\right) $ there is a natural action on $\mathcal{T}%
_{g,N_{0}}(L)$ of the mapping class group $\mathcal{M}ap_{g,N_{0}}$ defined
by the group of all the isotopy classes of orientation preserving
homeomorphisms of $\Omega $ which leave each boundary component $\partial
\Omega _{j}$ pointwise (and isotopy-wise) fixed. This action changes the
marking $f$ of \ $S_{g,N_{0}}$ on $\Omega $, and characterizes the quotient
space
\begin{equation}
\mathcal{M}_{g,N_{0}}(L)\doteq \frac{\mathcal{T}_{g,N_{0}}(L)}{\mathcal{M}%
ap_{g,N_{0}}}
\end{equation}
as the moduli space of Riemann surfaces (homeomorphic to $S_{g,N_{0}}$) with 
$N_{0}$ boundary components of length $|\partial \Omega _{j}|=L_{j}$. Note
again that when  $\left\{ L_{j}\rightarrow 0\right\} $, $\mathcal{M}%
_{g,N_{0}}(L)$ reduces to the usual moduli space $\mathcal{M}_{g,N_{0}}$ of
Riemann surfaces of genus $g$ with $N_{0}$ punctures. We have $dim_{\mathbb{R}}\mathcal{M}%
_{g,N_{0}}(L)$ $=$ $6g-6+3N_{0}$ and $dim_{\mathbb{R}}\mathcal{M}%
_{g,N_{0}}$ $=$ $6g-6+2N_{0}$ the extra $N_{0}$ coming from the boundary lengths.
Let us denote by $\overline{\mathcal{M}}_{g,N_{0}}$\ the

Deligne-Mumford compactification  of the moduli space of \ $N_{0}$-pointed\ closed surfaces
of genus $g$. As a connected complex obifold  $\overline{\mathcal{M}}_{g,N_{0}}$ is naturally
endowed with the i-th tautological line
bundle $\mathcal{L}(i)$
whose fiber at the point $(\Omega ,p_{1},...,p_{N_{0}})\in \overline{%
\mathcal{M}}_{g,N_{0}}$ is the cotangent space of $\Omega $ at $p_{i}$.\, Let us recall also that
for surfaces with punctures $\in \mathcal{T}_{g,N_{0}}$ one can introduce a
trivial bundle, Penner's decorated Teichm\"{u}ller space, 
\begin{equation}
\widetilde{\mathcal{T}}_{g,N_{0}}\overset{\pi _{hor}}{\longrightarrow }%
\mathcal{T}_{g,N_{0}}
\end{equation}
whose fiber over a punctured surface $\widetilde{\Omega }$ is the set of all 
$N_{0}$-tuples of horocycles in $\widetilde{\Omega }$, with one horocycle
around each puncture, (there is a corresponding trivial fibration $%
\widetilde{\mathcal{M}}_{g,N_{0}}$ over $\mathcal{M}_{g,N_{0}}$). A section
of this fibration is defined by choosing the total length of the horocycle
assigned to each puncture in $\widetilde{\Omega }$. 

\medskip

\noindent In our case, the hyperbolic surfaces $\Omega $ with
geodesic boundary  $\in \mathcal{T}_{g,N_{0}}(L)$ arise from the
interplay between the geometry of the horocycles  and
the conical holonomies $e^{i\sigma _{3}\Theta (k)}$ of the underlying Regge
triangulation $|T_{l}|\rightarrow M$. In particular, if let $%
\left\{ \Theta (k)\rightarrow 2\pi \right\} $, then there is a natural  mapping
between $(\mathbb{L}^{+})^{N_{0}}\times \mathcal{T}_{g,N_{0}}(L)$, ($\mathbb{%
L}^{+}$ being the future light cone parametrizing the horospheres), and the
decorated Teichm\"{u}ller space $\widetilde{\mathcal{T}}_{g,N_{0}}$. This \emph{conical forgetful} mapping is defined by associating to $\Omega $  the hyperbolic surface $\widetilde{\Omega }$\ with $N_{0}$
punctures obtained by letting $\left\{ \Theta (k)\rightarrow 2\pi \right\} $
and by decorating the resulting cusps with the horocycles traced by the horospheres $\left\{ \Sigma
_{k}\right\} $, \emph{i.e.}
\begin{gather}
(\mathbb{L}^{+})^{N_{0}}\times \mathcal{T}_{g,N_{0}}(L)\longrightarrow 
\widetilde{\mathcal{T}}_{g,N_{0}} \label{forgetcone} \\
\left( \left\{ \Sigma _{k}\right\} ,\Omega \cup \left\{ \sqcup
_{k=1}^{N_{0}(T)}(\partial \Omega _{k},|\ln \frac{\Theta (k)}{2\pi }%
|\right\} \right) \longmapsto \widetilde{\Omega }\doteq \left( \left\{
\Sigma _{k}\right\} ,\Omega \simeq M\setminus V(T)\right) ,  \notag
\end{gather}
where $V(T)$ denotes the set of \ $N_{0}(T)$  vertex of $M$. 
 In such a construction an interesting role is played by the
lambda-lengths associated with the decorated edges of the triangles $\sigma
_{hyp}^{2}$. According to a classical result by Penner (\cite{Penner2}, Theorem
3.3.6), the pull-back $\pi _{hor}^{\ast }\omega _{WP}$ under the map $\pi
_{hor}:\widetilde{\mathcal{T}}\mathcal{\,}_{g}^{s}\rightarrow \mathcal{T}%
_{g}^{s}$ of the Weil-Petersson K\"{a}hler two-form $\omega _{WP}$ is given
by 
\begin{equation}
-2\sum_{[\sigma _{hyp}^{2}]\,\,}d\,\ln
\,\lambda _{0}\,\wedge \,d\,\ln \,\lambda _{1}+\,d\,\ln \,\lambda
_{1}\,\wedge \,d\,\ln \,\lambda _{2}+\,d\,\ln \,\lambda _{2}\,\wedge
\,d\,\ln \,\lambda _{0}\,,  \label{WeiPet}
\end{equation}
where the sum runs over all ideal triangles $\sigma _{hyp}^{2}$ whose
ordered edges take the lambda-lengths $\lambda _{0}$, $\lambda _{1}$, $%
\lambda _{2}$. Note that (for dimensional reason) $\pi ^{\ast }\omega _{WP}$
is a degenerate pre-symplectic form. Either by pulling back $\pi
_{hor}^{\ast }\omega _{WP}$ one more time under the action of the conical
forgetful mapping (\ref{forgetcone}), or by analyzing its invariance properties under the
mapping class group, is straightforward to verify that (\ref{WeiPet}) 
extends also to bordered case. In our setting  it provides 
\begin{gather}
\pi _{hor}^{\ast }\omega _{WP}(\Sigma )=  \label{EucWP} \\
-2\sum_{_{[\sigma _{hyp}^{2}]_{F(T)}\,}\,\,}d\,\ln \,\lambda \left( \Sigma
_{k},\Sigma _{h}\right) \,\wedge \,d\,\ln \,\lambda \left( \Sigma
_{h},\Sigma _{j}\right) +\,  \notag \\
+d\,\ln \,\lambda \left( \Sigma _{h},\Sigma _{j}\right) \,\wedge \,d\,\ln
\,\lambda \left( \Sigma _{j},\Sigma _{k}\right) +d\,\ln \,\lambda \left(
\Sigma _{j},\Sigma _{k}\right) \,\wedge \,d\,\ln \,\lambda \left( \Sigma
_{k},\Sigma _{h}\right) =  \notag \\
\notag \\
=\,-2\sum_{F(t)\,\,}\frac{d\,l(k,h)\wedge d\,l(h,j)}{l(k,h)l(h,j)}+\frac{%
d\,l(h,j)\wedge d\,l(j,k)}{l(h,j)l(j,k)}+\frac{d\,l(j,k)\wedge d\,l(k,h)}{%
l(j,k)l(k,h)}\,,  \notag
\end{gather}
where we have exploited the expressions (\ref{luno}), (\ref{ldue}), (\ref{ltre}) providing the lambda-lengths in terms of
the Euclidean edge-lengths $l(k,h)$ of the Regge triangulation. 

\medskip

\noindent If we denote by  $V_{g,N_{0}}\left( \mathcal{M}_{g,N_{0}}(L)\right) $ the
volume of the moduli space $\mathcal{M}_{g,N_{0}}(L)$ with respect to the
measure associated with the Weil-Petersson form $\omega _{WP}(\Sigma )$,
then one can compute the dependence of \ $V_{g,N_{0}}\left( \mathcal{M}%
_{g,N_{0}}(L)\right) $ from the boundary lengths $L=\left(
L_{1},...,L_{N_{0}}\right) $\  by exploiting a remarkable result due to M.
Mirzakhani \cite{mirzakhani1, mirzakhani2} 
\begin{theorem}
(Maryam Mirzakhani (2003))\\
 \noindent The Weil-Petersson volume $V_{g,N_{0}}\left( \mathcal{M}%
_{g,N_{0}}(L)\right) $ is a polynomial in $L_{1},...,L_{N_{0}}$%
\begin{equation}
V_{g,N_{0}}\left( \mathcal{M}_{g,N_{0}}(L)\right) =\sum_{\underset{|\alpha
|\leq 3g-3+N_{0}}{(\alpha _{1},...,\alpha _{N_{0}})\in (\mathbb{Z}_{\geq
0})^{N_{0}}}}C_{\alpha _{1}...\alpha _{N_{0}}}L_{1}^{2\alpha
_{1}}...L_{N_{0}}^{2\alpha _{N_{0}}},
\end{equation}
where $|\alpha |=\sum_{i=1}^{N_{0}}\alpha _{i}$ and where the coefficients $%
C_{\alpha _{1}...\alpha _{N_{0}}}>0$ are (recursively determined) numbers of the form 
\begin{equation}
C_{\alpha _{1}...\alpha _{N_{0}}}=\pi ^{6g-6+2N_{0}-2|\alpha |}\cdot q
\end{equation}
for  rationals  $q\in \mathbb{Q}$.
\end{theorem}
\ Moreover \ Mirzakhani \cite{mirzakhani1, mirzakhani2} is
also able \ to express $\ C_{\alpha _{1}...\alpha _{N_{0}}}\ \ $in terms of
the intersection numbers $<\tau _{\alpha _{1}}...\tau _{\alpha _{N_{0}}}>$ \cite{witten} of the tautological line bundles  $\mathcal{L}(i)$  over
$\overline{\mathcal{M}}_{g,N_{0}}$\,
according to
\begin{equation}
C_{\alpha _{1}...\alpha _{N_{0}}}=\frac{2^{m(g,N_{0})|\alpha |}}{2^{|\alpha |}\prod_{i=1}^{N_{0}}%
\alpha _{i}!(3g-3+N_{0}-|\alpha |)!}<\tau _{\alpha _{1}}...\tau _{\alpha _{N_{0}}}>,
\end{equation}
\begin{equation}
<\tau _{\alpha _{1}}...\tau _{\alpha _{N_{0}}}>\doteq \int_{\overline{\mathcal{M}}_{g,N_{0}}}\psi _{1}^{\alpha
_{1}}...\psi _{N_{0}}^{\alpha _{N_{0}}}\cdot \omega _{WP}^{3g-3+N_{0}-|\alpha |}
\end{equation}
where $\psi _{i}$ is the first Chern class of $\mathcal{L}(i)$, and where $m(g,N_{0})\doteq \delta _{g,1}\delta _{N_{0},1}$.
Note in particular that the constant term $C_{0...0}$\ of the polynomial $%
V_{g,N_{0}}\left( \mathcal{M}_{g,N_{0}}(L)\right) $ is the volume of  $\overline{\mathcal{M}}_{g,N_{0}}$\  \emph{i.e.},
\begin{equation}
C_{0...0}=V_{g,N_{0}}\left( \overline{\mathcal{M}}_{g,N_{0}}\right) =\int_{%
\overline{\mathfrak{M}}_{g},_{N_{0}}}\frac{\omega _{WP}{}^{3g-3+N_{0}(T)}}{%
\left( 3g-3+N_{0}(T)\right) !} \label{wpvol} \\
\end{equation}
The valutation of the $C_{\alpha _{1}...\alpha _{N_{0}}}$ is at fixed $N_{0}$  (and at fixed genus $g$), and it is interesting to compare it to what is known when the genus increases or when $N_{0}$ increases. In particular, let
us recall that the Weil-Petersson volume of the moduli space $\overline{\mathcal{M}}_{g,N_{0}}$ for any fixed value of \ $N_{0}$\ is such that 
\begin{equation}
A_{1}^{g}(2g)!\leq V_{g,N_{0}}\left( \overline{\mathcal{M}}_{g,N_{0}}\right)\leq
A_{2}^{g}(2g)!,
\end{equation}
where the constants $0<A_{1}<A_{2}$ are independent of $N_{0}$ (see \cite{schumacher, grushevsky}).
Conversely,
the large $N_{0}$ asymptotics of $%
V_{g,N_{0}}\left( \overline{\mathcal{M}}_{g,N_{0}}\right)$ at fixed genus has been discussed by Manin and
Zograf \cite{zograf, manin}. They obtained the asymptotic series
\begin{gather}
V_{g,N_{0}}\left( \overline{\mathcal{M}}_{g,N_{0}}\right)=\pi ^{6g-6+2N_{0}}\times
\label{manzog} \\
\times (N_{0}+1)^{\frac{5g-7}{2}}C^{-N_{0}}\left( B_{g}+\sum_{k=1}^{\infty }%
\frac{B_{g,k}}{(N_{0}+1)^{k}}\right) ,  \notag
\end{gather}
where $C=-\frac{1}{2}j_{0}\frac{d}{dz}J_{0}(z)|_{z=j_{0}}$, ($J_{0}(z)$\ the
Bessel function, $j_{0}$ its first positive zero); (note that $C\simeq
0.625....$). The genus dependent parameters $B_{g}$\ are explicitly given
\cite{manin} by 
\begin{equation}
\left\{ 
\begin{tabular}{ll}
$B_{0}=\frac{1}{A^{1/2}\Gamma (-\frac{1}{2})C^{1/2}},$ & $B_{1}=\frac{1}{48}%
, $ \\ 
$B_{g}=\frac{A^{\frac{g-1}{2}}}{2^{2g-2}(3g-3)!\Gamma (\frac{5g-5}{2})C^{%
\frac{5g-5}{2}}}\left\langle \tau _{2}^{3g-3}\right\rangle ,$ & $g\geq 2$%
\end{tabular}
\right.
\end{equation}
where $A\doteq -j_{0}^{-1}J_{0}^{\prime }(j_{0})$, and $\left\langle \tau
_{2}^{3g-3}\right\rangle $ is a Kontsevich-Witten \cite{kontsevich} intersection number, (the
coefficients $B_{g,k}$ can be computed similarly-see \cite{manin} for details).

\subsection{\protect\bigskip An example of open/closed string duality}

The preceding results provide a suitable kinematical set up for establishing a open/closed string duality once the appropriate field decoration is activated. To this end let us consider the non-dynamical null twistors fields decorating the vertex of the Regge triangulation. These fields geometrically describe geodesics in $\mathbb{H}^{3}$, with an end point at 
$\infty\in \partial \mathbb{H}^{3,+}_{up}$ $=$ $(\mathbb{R}^{2}\times \{0\})\cup\{\infty \}$,  projecting to the $N$ components $\partial \Omega _{k}$ of the  boundary of $\Omega $. Thus, they can be interpreted as fields on  $\Omega $ with preassigned Dirichlet conditions on the various boundary components $\partial \Omega _{k}$, and we can consider the N-point function on $\mathcal{M}_{g,N_0}(L)$, describing correlations between such Dirichlet conditions. Explicitly,
let us consider the
$\lambda $-lengths $\lambda (\Sigma _{\infty },\Sigma _{k})$ associated with the
vertical geodesic connecting $v^{0}(0)\simeq \infty $ with the\ \ generic
vertex $v^{0}(k)$ of the ideal triangulation $H(|T|\rightarrow M)$. We form
the expression
\begin{gather}
Z^{open}_{N_{0} ,g}\left((L_{1},\delta (\Sigma _{\infty },\Sigma _{1});...;(L_{N_{0}},\delta (\Sigma
_{\infty },\Sigma _{N_{0}})\right) \doteq  \\
=\sum_{\underset{|\alpha |\leq 3g-3+N_{0}}{(\alpha _{1},...,\alpha
_{N_{0}})\in (\mathbb{Z}_{\geq 0})^{N_{0}}}}C_{\alpha _{1}...\alpha _{N_{0}}}%
\left[ \lambda (\Sigma _{\infty },\Sigma _{1})\,L_{1}\right] ^{2\alpha
_{1}}...\left[ \lambda (\Sigma _{\infty },\Sigma _{N_{0}})\,L_{N_{0}}\right]
^{2\alpha _{N_{0}}}=  \notag \\
=2^{\left| \alpha \right| }\sum_{\underset{|\alpha |\leq 3g-3+N_{0}}{(\alpha
_{1},...,\alpha _{N_{0}})\in (\mathbb{Z}_{\geq 0})^{N_{0}}}}C_{\alpha
_{1}...\alpha _{N_{0}}}\,e^{\alpha _{1}\,\delta (\Sigma _{\infty },\Sigma
_{1})}\,L_{1}^{2\alpha _{1}}...e^{\alpha _{N_{0}}\,\delta (\Sigma _{\infty
},\Sigma _{N_{0}})}\,\,L_{N_{0}}^{2\alpha _{N_{0}}},  \notag
\end{gather}
where  $\delta (\Sigma _{\infty },\Sigma _{k})$ is the signed hyperbolic
distance between the respective horosphere. \ Thus, $Z^{open}_{N_{0} ,g}\left(
;\delta (\Sigma _{\infty },\Sigma _{1}),...,\right) $ basically provides  correlations in the moduli space $\overline{\mathcal{M}}_{g,N_{0}}(L)$ among the  Dirichlet boundary
conditions, along the $\left\{ \partial \Omega _{k}\right\} $ boundary
components, of the local fields $\delta (\Sigma _{\infty },\Sigma _{k})$. Such correlations describe the distribution in
$\overline{\mathcal{M}}_{g,N_{0}}(L)$ of the (hyperbolic) distance from the
surfaces $\Omega \in \overline{\mathcal{M}}_{g,N_{0}}(L)$ and the Euclidean screen  $%
\Sigma _{\infty }\subset \mathbb{H}_{hyp}^{3}$ from which the generic
hyperbolic surface $\Omega $ is locally generated by projecting Euclidean triangles into hyperbolic triangles.  
From  Mirzakhani's results we get
\begin{gather}
Z^{open}_{N_{0} ,g}\left((L_{1},\delta (\Sigma _{\infty },\Sigma _{1});...\right) = \frac{1}{(3g-3+N_{0}-|\alpha |)!}\,\times \\
\times \sum_{\underset{|\alpha |\leq 3g-3+N_{0}}{%
(\alpha _{1},...,\alpha _{N_{0}})\in (\mathbb{Z}_{\geq 0})^{N_{0}}}}\int_{%
\overline{\mathcal{M}}_{g,N_{0}}}\prod_{i=1}^{N_{0}}\frac{L_{i} ^{2\alpha _{i}}e^{\alpha _{i}\,\delta (\Sigma
_{\infty },\Sigma _{i})}\,}{\alpha _{i}!}\psi _{i}^{\alpha _{i}}\,\omega
_{W\,P}^{3g-3+N_{0}-|\alpha |}.  \notag
\end{gather}
We could insert here the explicit expression of the boundary lengths $\,L_{i}$ $=$ $\left( \ln \frac{%
\Theta (i)}{2\pi }\right)$, however, for our purposes
it is more  interesting to consider the scaling regime in which  the projection
field $\left\{ \delta (\Sigma _{\infty },\Sigma _{k})\right\} $ \ generates the bordered hyperbolic surface $\Omega $ from
the $N_{0}$-pointed closed  surface $M\setminus V(T)$ associated with the random Regge triangulation $|T_{l}|\rightarrow M$. According to formula (31) governing the distance scaling in hyperbolic three-geometry,  such a regime corresponds all possible  rescalings $\delta (\Sigma _{\infty },\Sigma
_{k})\rightarrow \beta \,\delta (\Sigma _{\infty },\Sigma
_{k}) $ and $L_{k}\rightarrow \beta^{-1}\, L_{k}$,\, $\beta \in (0,\infty )$, of 
the hyperbolic distance  and the boundary lengths,  
such that
\begin{equation}
L_{k}(\beta )
e^{\delta_{\beta } (\Sigma _{\infty },\Sigma _{k})}= \left| \ln \frac{%
\Theta (k)}{2\pi }\right|\doteq    t_{k}^{\frac{1}{2}},\,\,\,\beta \in (0,\infty ).
\end{equation}
Under such regime, we get 
\begin{gather}
Z^{open}_{N_{0} ,g}\left((L_{1}(\beta ),\delta_{\beta } (\Sigma _{\infty },\Sigma _{1});...\right) 
= Z^{closed}_{N_{0} ,g}\left(t_{1}...t_{N_{0}}\right) \label{opcl} \\
 \doteq \frac{1}{%
(3g-3+N_{0}-|\alpha |)!}\sum_{\underset{|\alpha |\leq 3g-3+N_{0}}{(\alpha
_{1},...,\alpha _{N_{0}})\in (\mathbb{Z}_{\geq 0})^{N_{0}}}}\int_{\overline{%
\mathcal{M}}_{g,N_{0}}}\prod_{i=1}^{N_{0}}\frac{t_{i}^{\alpha _{i}}\,}{%
\alpha _{i}!}\psi _{i}^{\alpha _{i}}\,\omega _{W\,P}^{3g-3+N_{0}-|\alpha |},
\notag
\end{gather}
\newline
which is the generating function, at finite  $N_{0}$ and at finite genus $g$%
, of the intersection theory \cite{witten, looijenga} over the moduli space $\overline{\mathcal{M}}%
_{g,N_{0}}$ of closed \ $N_{0}$-pointed genus $g$ surfaces $M\setminus V(T)$ associated with random Regge triangulations
$|T_{l}|\rightarrow M$. Note that the Weil-Peterson form  $\omega _{W\,P}$ in (\ref{opcl}) can be appropriately interpreted in the simplicial form (\ref{EucWP}).  The open/closed surface duality mapping (\ref{opcl})
extends to the moduli spaces $\overline{\mathcal{M}}_{g,N_{0}}(L)$ and $%
\overline{\mathcal{M}}_{g,N_{0}}$  the geometric duality between the
hyperbolic (ideally triangulated surface) with geodesic boundary $\Omega $ $=
$  $H(|T|\rightarrow M)$ and the closed $N_{0}$-pointed surface $%
\widetilde{\Omega }$ associated with a random Regge triangulation. 
Note that the duality (\ref{opcl}) can be immediately rephrased in twistorial terms by recalling
the connection (\ref{spinnorm}) between the $\lambda $-lengths and the associated null
twistors in $\mathbb{C}P^{1}\times \mathbb{C}P^{1}\backslash \,\Delta $ \
given by
\begin{equation}
\lambda (\Sigma _{\infty },\Sigma _{k})=\frac{1}{\sqrt{2}}\left\| \xi
_{B}(\infty )\xi ^{B}(k)\right\| ,
\end{equation}
where $W^{\Lambda }(\infty ,k)\doteq (\xi ^{A}(\infty ),\overline{\xi }%
_{A^{\prime }}(k))$ is the null twistor corresponding to the unique geodetic
in $\mathbb{H}^{3}$ connecting the two horospheres $\Sigma _{\infty }$ and 
$\Sigma _{k}$. We get 
\begin{equation}
Z_{N_{0},g}^{\,open}((L_{1},\delta (\Sigma _{\infty },\Sigma
_{1}));...)=Z_{N_{0},g}^{\,open}((L_{1},W^{\Lambda }(\infty ,k));...),
\end{equation}
where
\begin{gather}
Z_{N_{0},g}^{\,open}((L_{1},W^{\Lambda }(\infty ,k));...)= \frac{1}{2^{2|\alpha |}(3g-3+N_{0}-|\alpha |)!}\times \\
\times \sum_{\underset{|\alpha |\leq
3g-3+N_{0}}{(\alpha _{i})\in (\mathbb{Z}_{\geq 0})^{N_{0}}}}\int_{\overline{%
\mathcal{M}}_{g,N_{0}}}\prod_{i=1}^{N_{0}}\frac{L_{i}^{2\alpha
_{i}}\,\left\| \xi _{B}(\infty )\xi ^{B}(i)\right\| ^{2\alpha _{i}}}{\alpha
_{i}!}\psi _{i}^{\alpha _{i}}\,\omega _{W\,P}^{3g-3+N_{0}-|\alpha |},  \notag
\end{gather}
and the open/closed duality can be interpreted in terms of the scaling
behavior of \linebreak $L_{i}^{2\alpha _{i}}\,\left\| \xi _{B}(\infty )\xi
^{B}(i)\right\| ^{2\alpha _{i}}$, \emph{i.e.}, of \  twistorial field
insertions at the vertices of the closed pointed surface associated with $%
|T_{l}|\rightarrow M$. 

\medskip

\noindent In order to describe a full-fledged open/closed
string duality in such a setting let us recall that,  according to the
 discussion in the preceding paragraph, the hyperbolic bordered surface $\Omega $\ carries the
gauge degrees of freedom associated with the flat $SU(2)$ connection $A_{(k)}
$ defined by (\ref{gpoten}). We can promote such $A_{(k)}$ to be dynamical fields and consider the amplitude of the corresponding $SU(2)$\ \
Yang-Mills theory. For a surface $\Omega $ of genus $g$, area \ $A(\Omega )$%
, and $N_{0}(T)$ geodesic boundary components $\left\{ \partial \Omega
_{k}\right\} $ with given holonomies $\left\{ U_{k}=e^{i\sigma _{3}\Theta
(k)}\right\} $, such an amplitude is \ given by the celebrated formula (see \emph{e.g.}, \cite{Moore}) 
\begin{gather}
Z^{Y-M}\left( U_{1},...,U_{No};A(\Omega )\right) = \label{zetaYM} \\
=\sum_{R\in \left\{
Unitary\;Irreps\right\} }\left( \dim \,R\right) ^{2-2g-N_{0}}\,e^{-\beta
_{YM}^{2}\,A(\Omega )\,C_{2}(R)}\prod_{i=1}^{N_{0}}\chi _{R}(U_{i})\notag,
\end{gather}
where the sum runs over all unitary irreducible representations of $SU(2)$, $%
\chi _{R}(U_{i})$ are the associated characters, $C_{2}(R)$ is the
eigenvalue of the quadratic Casimir in the representation $R$, and finally, $%
\beta _{YM}^{2}$ is the gauge coupling. Since we are considering a
hyperbolic surface with geodesic boundary, by the Gauss-Bonnet theorem 
\begin{equation}
\int_{\Omega }K_{\Omega }\,dA+\int_{\partial \Omega }k_{\partial \Omega
}\,ds=2\pi \,(2-2g-N_{0}),
\end{equation}
(where $K_{\Omega }=-1$ and $k_{\partial \Omega }=0$ are the Gaussian
curvature of $\Omega $ and the geodesic curvature of $\partial \Omega
=\sqcup \partial \Omega _{k}$, respectively), we get 
\begin{equation}
A(\Omega )=2\pi \,(2g-2+N_{0}).
\end{equation}
From this, and specializing (\ref{zetaYM}) to the irreps of $SU(2)$, we
eventually get 
\begin{gather}
Z^{Y-M}\left( U_{1},...,U_{No};A(\Omega )\right) = \\
=\sum_{j\in \left\{ \frac{1%
}{2}\mathbb{Z}_{+}\right\} }\left( \frac{e^{-\,2\pi \,\,j(j+1)\beta
_{YM}^{2}}}{2j+1}\right) ^{(2g-2+N_{0})}\prod_{i=1}^{N_{0}}\frac{\sin \left[
(2j+1)\frac{\Theta (i)}{2}\right] }{\sin \frac{\Theta (i)}{2}} \notag.
\end{gather}
In line with the preceding analysis, it is natural to consider the full
partition (actually a \ $N_{0}$-points)\ \ function  
\begin{gather}
Z_{N_{0},g,\beta _{YM}^{2}}^{\,open}((L_{1},W^{\Lambda }(\infty ,k),\Theta
(k));...)\doteq \frac{1}{2^{2|\alpha |}(3g-3+N_{0}-|\alpha |)!}\times  \\
\times \sum_{\underset{|\alpha
|\leq 3g-3+N_{0}}{(\alpha _{i})\in (\mathbb{Z}_{\geq 0})^{N_{0}}}}\sum_{j\in
\left\{ \frac{1}{2}\mathbb{Z}_{+}\right\} }\left( \frac{e^{-\,2\pi
\,\,j(j+1)\beta _{YM}^{2}}}{2j+1}\right) ^{(2g-2+N_{0})}\times   \notag \\
\times \int_{\overline{\mathcal{M}}_{g,N_{0}}}\prod_{i=1}^{N_{0}}\frac{%
L_{i}^{2\alpha _{i}}\,\left\| \xi _{B}(\infty )\xi ^{B}(i)\right\| ^{2\alpha
_{i}}}{\alpha _{i}!}\frac{\sin \left[ (2j+1)\frac{\Theta (i)}{2}\right] }{%
\sin \frac{\Theta (i)}{2}}\psi _{i}^{\alpha _{i}}\,\omega
_{W\,P}^{3g-3+N_{0}-|\alpha |},  \notag
\end{gather}
and discuss under what conditions it yields a duality of the form (\ref{opcl}). The
analysis of this problem is quite more complicated than the one discussed so far. It goes far beyond the pure kinematical aspects analized here since it involves an in depth analysis of the boundary conformal field theory associated with the modular dynamics of the  SU(2) gauge field on the bordered surface $\Omega $. It will be presented in a companion paper which is in preparation. However, already at this stage  it is clear that we are calling into play also the moduli spaces of stable bundles, in particular the variety $Hom(\pi _{1}(\Omega ),SU(2))/SU(2)$
of representations, up to conjugacy, of the fundamental group of the bordered surface $\Omega $ in $SU(2)$ such that the monodromy around $\partial \Omega _{k}$ lies in the conjugacy class of $\partial \Omega _{k}$. It is well known that such a representation variety can be identified with the moduli space of semi-stable holomorphic rank 2 vector
bundles over $\Omega $. In such a framework, fixing the conjugacy
class of the monodromy around a boundary component $\partial \Omega _{k}$ plays the same role played by the datum of
the length of the geodesic boundary of $\partial \Omega _{k}$ in the case of the moduli space ${\mathcal{M}}_{g,N_{0}}(L)$. Thus, one is expecting that also intersection theory over $Hom(\pi _{1}(\Omega ),SU(2))/SU(2)$ plays a role in establishing open/closed duality, (in her analysis of the volume of ${\mathcal{M}}_{g,N_{0}}(L)$, Mirzhakani draws similar conclusions). Moreover, since the analysis of the geometry of $Hom(\pi _{1}(\Omega ),SU(2))/SU(2)$ and of boundary conformal field theory for SU(2) is intimately connected with Chern-Simons theory, one may wonder if there is an explicit geometrical counterpart of this, analogous to the simple open/closed duality between hyperbolic bordered surfaces and random Regge triangulations. Quite remarkably, this is indeed the case, and we conclude our kinematical analysis of open/closed string duality by presenting, in the next section, the geometrical aspects involved in activating Chern-Simons theory. 

\section{\protect\bigskip Connection with Hyperbolic 3-manifold}
 
The connection between twistorially decorated Regge surfaces and hyperbolic surfaces with boundaries can be naturally extended to three-dimensional hyperbolic cone-manifolds. Recall that to the twistor field $\sigma ^{0}(k)$ $\longmapsto$ $W^{\Lambda }(k,\infty )$  on $|T_{l}|\rightarrow M$ we can associate either the marked horosphere $(\Sigma _{k},z_{k})$ or, equivalently, the (unique) geodesic $\gamma(k,\infty ) $ in $\mathbb{H}_{up}^{\,3,+}$  connecting the vertex $v^{0}(k)$ with the vertex at $\infty $ of an ideal
tetrahedron $\sigma _{hyp}^{3}(\infty ,k,h,j)$ in $\mathbb{H}_{up}^{\,3,+}$. 
In particular, to any two adjacent\ triangles sharing a common edge, say \ $%
\sigma ^{2}(k,h,j)$ and $\sigma ^{2}(k,j,l)$, correspond pairwise adjacent
tetrahedra, $\sigma _{hyp}^{3}(\infty ,k,h,j)$ and $\sigma _{hyp}^{3}(\infty
,k,j,l)$, that can be glued along the isometric faces $\sigma _{hyp}^{2}(j,k,\infty )$
and $\sigma_{hyp} ^{2}(\infty ,k,j)$.
Each face-pairing is realized by an isometry of  $\mathbb{H}_{up}^{3,+}$ 
\begin{equation}
f_{jk}:\sigma _{hyp}^{2}(j,k,\infty )\longrightarrow \sigma
_{hyp}^{2}(\infty ,k,j)
\end{equation}
which reverses orientation (so as to have orientability of the
resulting complex). In this way, by pairwise glueing  the $q(k)$ ideal
tetrahedra $\left\{ \sigma _{hyp}^{3}(\infty ,k,h_{\alpha },h_{\alpha +1})\right\} $, associated with the corresponding Euclidean triangles $%
\sigma ^{2}(k,h_{\alpha },h_{\alpha +1})$, we generate
a polytope
\begin{equation}
P^{3}(k) \doteq \left.
\coprod_{\alpha =1}^{q(k)}\sigma _{hyp}^{3}(\infty ,k,h_{\alpha },h_{\alpha
+1})\right/ \left\{ f_{h_{\alpha }k}\right\}
\label{ktope} 
\end{equation}
with a conical singularity along the core geodesic 
$\gamma (k,\infty )$. Explicitly, let us denote by $\psi _{\cdot   ,\cdot  }$ the dihedral
angles associated with the edges $\sigma _{hyp}^{1}(\cdot ,\cdot )$ \ of
this polytope. From the relations between the dihedral angles of each hyperbolic
tetrahedron $\sigma _{hyp}^{3}(\infty ,k,h,j)$ and the vertex angles of the
corresponding Euclidean triangle $\sigma ^{2}(k,h,j)$ it easily follows that 
\begin{gather}
\psi _{\infty ,\,h}=\theta _{hj\,k}+\theta _{k\,jl}\;, \\
\psi _{k\,j}=\theta _{khj}+\theta _{klj}\;,  \notag \\
\psi _{hj}=\theta _{hkj}\;,  \notag \\
\psi _{\infty ,\,k}=\sum_{\alpha =1}^{q(k)}\theta _{\alpha ,\,k,\,\alpha +1}=\Theta (k).
\notag
\end{gather}
Note in particular that the conical defect $\Theta (k)$ at the vertex $\sigma ^{0}(k)\in
Star[\sigma ^{0}(k)]$ propagates as a conical defect along the core geodesic $%
\gamma (k,\infty )$ of $\mathbb{H}_{up}^{3,+}$. It follows that $P^{3}(k)$ has a non-complete hyperbolic metric and that the singularity on $\gamma (k,\infty )$ is conical with angle $\Theta (k)$.  
In order to endow $P^{3}(k)$ with a hyperbolic structure, let $\widetilde{P}_{\gamma }(k)$ denote the universal cover in $%
\mathbb{H}_{up}^{3,+}$ of \ $P^{3}(k)$, with the core geodesic $\gamma
(k,\infty )$ removed. $\widetilde{P}_{\gamma }(k)$ carries a natural
hyperbolic structure and the holonomy representation of its fundamental group, $\pi _{1}(\widetilde{P}%
_{\gamma }(k))=\mathbb{Z}$, is generated by an isometry of $\ \widetilde{P}%
_{\gamma }(k)\subset \mathbb{H}_{up}^{3,+}$ of the form
\begin{gather}
\rho(k) :\pi _{1}(\widetilde{P}_{\gamma }(k))\longrightarrow Isom\left( \mathbb{%
H}_{up}^{3,+}\right) \label{isohyp}  \\
(c_{s},s)\longmapsto \left[ a(k)\left( 
\begin{tabular}{ll}
$e^{i\,\phi (s)}$ & $0$ \\ 
$0$ & $e^{-i\,\phi (s)}$%
\end{tabular}
\right) ,s\right] ,  \notag
\end{gather}
where $a(k)>1$ and $s\mapsto c_{s}$, $0\leq s<\infty $ is \ closed curve
winding around the link of \ $\sigma ^{0}(k)$ in  $Star\left[ \sigma ^{0}(k)%
\right] $ with $\phi (s=2\pi )=\Theta (k)$. Since  $a(k)>1$, the isometry is hyperbolic
(fixing the point $v^{0}(k)$ and $\infty $ in $\mathbb{H}_{up}^{3,+}$).
 For simplicity, let us
identify $v^{0}(k)$ with the origin of $\mathbb{H}_{up}^{3,+}$. The horosphere $\Sigma _{k}$ intersects $\sqcup \sigma _{hyp}^{3}(\infty
,k,h_{\alpha },h_{\alpha +1})$ along a sequence of offset horocycle segments 
$\left\{ \digamma _{k}^{h_{\alpha }}\right\} $ such that 
\begin{equation}
d_{\mathbb{H}^{3}}(\digamma _{k}^{h_{1}},\widehat{\digamma }%
_{k}^{h_{q(k)}})=\left| \ln \,\frac{\sum_{\alpha =1}^{q(k)}\theta _{\alpha
+1,k,\alpha }}{2\pi }\right| .
\end{equation}
Similarly the concentric horosphere $^{\ast }\Sigma _{k}$ defined by $%
z=a(k)\,z_{k}$, $a(k)\geq 1$, intersects the $\sqcup \sigma _{hyp}^{3}(\infty
,k,h_{\alpha },h_{\alpha +1})$ along a sequence of horocycle segments 
$\left\{ ^{\ast }\digamma _{k}^{h_{\alpha }}\right\} $ such that \linebreak $d_{\mathbb{H%
}^{3}}(^{\ast }\digamma _{k}^{h_{1}},^{\ast }\widehat{\digamma }%
_{k}^{h_{q(k)}})=\left| \ln \,\frac{\Theta (k)}{2\pi }\right| $. \ Let us consider the
rectangular parallelepiped labeled by the segments $\left(
\digamma _{k}^{h_{1}},\widehat{\digamma }_{k}^{h_{q(k)}};^{\ast }\digamma
_{k}^{h_{1}},^{\ast }\widehat{\digamma }_{k}^{h_{q(k)}}\right) $. A
straightforward application of (\ref{reldistance})
 \ provides the following relations
between the (hyperbolic) lengths of the sides of this parallelepiped
\begin{eqnarray}
\left| ^{\ast }\widehat{\digamma }_{k}^{h_{q(k)}}\right|  &=&e^{-d_{\mathbb{H%
}^{3}}(\Sigma _{k},^{\ast }\Sigma _{k})}\left| \widehat{\digamma }%
_{k}^{h_{q(k)}}\right| , \\
\left| ^{\ast }\digamma _{k}^{h_{1}}\right|  &=&e^{-d_{\mathbb{H}%
^{3}}(\Sigma _{k},^{\ast }\Sigma _{k})}\left| \digamma _{k}^{h_{1}}\right| ,
\notag
\end{eqnarray}
where $\left| ...\right| $  denotes the length of the corresponding
horocycle segment. Since 
\begin{equation}
d_{\mathbb{H}^{3}}(\Sigma _{k},^{\ast }\Sigma _{k})=2\,\tanh ^{-1}\frac{a(k)-1}{%
a(k)+1}=\ln \,a(k),
\end{equation}
we get 
\begin{equation}
\left| ^{\ast }\widehat{\digamma }_{k}^{h_{q(k)}}\right| =a(k)^{-1}\left| 
\widehat{\digamma }_{k}^{h_{q(k)}}\right| ,\;\;\left| ^{\ast }\digamma
_{k}^{h_{1}}\right| =a(k)^{-1}\left| \digamma _{k}^{h_{1}}\right| .
\end{equation}
Moreover, from (\ref{reenter}) we have
\begin{equation}
\left| \widehat{\digamma }_{k}^{h_{q(k)}}\right| =e^{\left| \ln \,\frac{%
\Theta (k)}{2\pi }\right| }\left| \digamma _{k}^{h_{1}}\right| ,\;\left|
^{\ast }\widehat{\digamma }_{k}^{h_{q(k)}}\right| =e^{\left| \ln \,\frac{%
\Theta (k)}{2\pi }\right| }\left| ^{\ast }\digamma _{k}^{h_{1}}\right| .
\end{equation}
By comparing these expressions, it follows that we can match the length of 
horocycle segment $^{\ast }\widehat{\digamma }_{k}^{h_{q(k)}}$ with the
length of the segment $\digamma _{k}^{h_{1}}$ if we choose the parameter $a(k)$
according to 
\begin{equation}
a(k)=e^{\left| \ln \,\frac{\Theta (k)}{2\pi }\right| }.
\end{equation}
Such a matching condition allows, under the action of \ $\rho(k) \left( \pi
_{1}\left( \widetilde{P}_{\gamma }(k)\right) \right) $, an (offset)
identification between opposite faces of $\left( \digamma
_{k}^{h_{1}},\widehat{\digamma }_{k}^{h_{q(k)}};^{\ast }\digamma
_{k}^{h_{1}},^{\ast }\widehat{\digamma }_{k}^{h_{q(k)}}\right) $, and consequently we can choose this rectangular parallelepiped
as a fundamental domain for the action of \ the holonomy representation $\rho(k) $. The resulting developing map
describes $\widetilde{P}_{\gamma }(k)$ as an incomplete manifold and $P_{hyp}^{3}(k)$ $\doteq $ $\widetilde{P}_{\gamma }(k)\setminus \rho(k) $ is
topologically equivalent to a solid torus $\mathbb{S}^{1}\times B^{2}$, ($B^{2}$ being the meridianal 2-dimensional disc) with
the central  geodesic missing. Note that such a geodesic can be naturally identified with the geodesic boundary component 
$\partial \Omega _{k}$ of the open hyperbolic surface $\Omega $. In order to get an intuitive picture of what happens, observe that the identification polytope $P^{3}(k)$, cut by the horosphere $\Sigma _{\infty }$, is topologically a solid cylinder sliced by the faces of the component tetrahedra. If we remove a tube of small (infinitesimal) width around the central geodesic $\gamma (k,\infty) $ we get a topological solid torus sliced into parallelepipeds, with a thin and long tubular hole associated with the removed geodesic. The isometry (\ref{isohyp}) twists up this solid torus with a shearing motion, like a 3-dimensional photographic diaphragm. Adjacent parallelepipeds slide one over the other tilting up, while the central tube correspondingly winds up accumulating towards an horizontal $\mathbb{S}^{1}$.

\subsection{\protect\bigskip Hyperbolic volume}
 
 We can formally
extend this geometric analysis to the whole Regge triangulation $\left|
T_{l}\right| \rightarrow M$ \ by forming the support space (for a compatible hyperbolic structure)
\begin{equation}
V\doteq \left. \coprod_{\sigma _{hyp}^{2}(l,m,\infty )}^{N_{1}(T)}\sigma
_{hyp}^{3}(\infty ,k,h,j)\right/ \left\{ f_{lm}\right\} ,
\label{Vmani}
\end{equation}
(the number of hyperbolic faces to be paired is equal to the
number $N_{1}(T)$ of edges in $\left| T_{l}\right| \rightarrow M$).
Note that the link of the vertex at $\infty $ in $V$ is  
\begin{equation}
link\,\,\left[ \infty \right] \doteq \bigcup_{\sigma
_{hyp}^{1}(l,m)}^{N_{1}(T)}\sigma _{hyp}^{2}(k,h,j),
\end{equation}
where the glueing along the edges $\left\{ \sigma _{hyp}^{1}(l,m)\right\} $
is modelled after the Regge surface $|T_{l}|\rightarrow M$. If this latter has genus 
$g$, then from the Euler  and  Dehn-Sommerville relations
\begin{gather}
N_{0}(T)-N_{1}(T)+N_{2}(T)=2-2g, \\
2N_{1}(T)=3N_{2}(T),  \notag
\end{gather}
we get that the support space $V$ has
\begin{equation}
N_{2}(T)=2N_{0}(T)+4g-4\geq N_{0}(T)+g
\end{equation}
\  ideal tetrahedra with $N_{0}(T)$ vertices associated with its boundary
components $\partial V$. As we have seen in section \ref{RTITS}, the edge-glueing of $\{\sigma _{hyp}^{2}(k,h,j)\}$ gives rise to an incomplete hyperbolic surface and
consequently also $V$ cannot support, as it stands, a complete hyperbolic structure. To take care of this, we start by removing from $V$ an open (horospherical) neighborhood of the vertices. In this way, each tetrahedron $\sigma
_{hyp}^{3}(\infty ,k,h,j)$ becomes a octahedron with four (Euclidean) triangular faces (in the same similarity class which defines the given tetrahedron), and four (hyperbolic) exagonal faces. Note that the boundary of the removed open neighborhood of $\infty$   is triangulated by Euclidean triangles and it reproduces $|T_{l}|\rightarrow M$. Note also that the removed neighborhoods cut out an open disk $D_{k}$ around each vertex $v^{0}(k)$ in $\partial V$.  Next, we remove from $V$ also an open neighborhood of the geodesics $\left\{ \gamma (k,\infty )\right\}_{k=1}^{N_{0}(T)}$. In this way we get from the support space $V$ a handlebody 
$H_{V}$. Topologically,  $H_{V}$ is $[0,1]\times \Omega$, where  $\Omega$ is the surface with boundary ($\sqcup\partial\Omega _{k} $) associated with the hyperbolic completion of  $\sqcup\sigma _{hyp}^{2}(k,h,j)$. The handlebody $H_{V}$ plays here the role of the polytope  $\widetilde{P}_{\gamma }(k)$ introduced in connection with the support space (\ref{ktope}). By identifying the bottom $\Omega _{0}\simeq  \partial H_{V}|_{0}$ and top $\Omega _{1}\simeq  \partial H_{V}|_{1}$ copies of the surface $\Omega$ by means of the appropriate orientation reversing boundary homeomorphism $h:\partial H_{V}|_{0}\rightarrow \partial H_{V}|_{1}$, with $h(\partial \Omega _{k}|_{0})$ $=$ $-\partial \Omega _{k}|_{1}$, we get the support space 
\begin{equation}
V\left( \left\{ \Theta (k)\right\} _{k=1}^{N_{0}(T)}  \right)\setminus K \doteq H_{V}\setminus \sim ^{h}
\end{equation}
($V\setminus K$, for notational ease), where $K$ is the knot-link generated in $H_{V}$ by the action of the identification homeomorphism $h$ on the boundaries connecting the tubes associated with the removed core geodesics $\left\{ \gamma (k,\infty )\right\}_{k=1}^{N_{0}(T)}$. It is  not yet obvious that $V\setminus K$ admits a complete hyperbolic structure. First,  
we have been rather cavalier on the delicate issue concerning orientation in glueing the ideal tetrahedra, (for semi-simplicial triangulations problems connected with orientability of the hyperbolic complexes obtained upon face-identifications can be rather serious and we may end up in a ideal triangulation which may actually not define a manifold). Moreover, around the removed geodesics $\left\{ \gamma (k,\infty )\right\}_{k=1}^{N_{0}(T)}$ the geometry is conical, and in order to establish completeness for the hyperbolic structure we have to discuss how hyperbolic Dehn filling can be extended to cone manifolds. These are delicate issues which, to the best of our knowledge, do not have answers that can be easily given in general terms. The interested reader may wish to consult the remarkable papers \cite{PetronioPorti, PetronioFrigerio, Moroianu} where particular cases are thoroughly discussed. Notwithstanding the technical difficulties in characterizing complete hyperbolic structures on $V\setminus K$, their existence, when established, implies a number of important consequences which bear relevance to our analysis.

\medskip 

\noindent First of all, if  the support space $V\setminus K$ generated by $\left| T_{l}\right| \rightarrow M$, is indeed a three-dimensional hyperbolic manifold $V_{hyp}(\left\{ \Theta (k)\right\}\setminus K$, then we can easily compute its hyperbolic volume in terms of the conical angles $(\left\{ \Theta (k)\right\} _{k=1}^{N_{0}(T)}) $. As a matter of fact we can associate to any
triangle $\sigma ^{2}(k,h,j)$ of $\left| T_{l}\right| \rightarrow M$ the volume $%
Vol[\sigma _{hyp}^{3}]$ of the corresponding ideal tetrahedron $\sigma
_{hyp}^{3}$. According to Milnor's formula, (see \emph{e.g.} \cite{benedetti}, for a very informative analysis), such a volume can be
expressed in terms of the Lobachevsky functions $\mathcal{L}(\theta
_{jk\,h}) $, $\mathcal{L}(\theta _{khj})$, and $\mathcal{L}(\theta _{hjk})$
of the respective vertex angles of $\sigma ^{2}(k,h,j)$, where 
\begin{equation}
\mathcal{L}(\theta _{jk\,h})\doteq -\int_{0}^{\theta _{jk\,h}}\ln \,\left|
2\sin \,x\right| \,dx.
\end{equation}
In our setting, this translates into the mapping 
\begin{gather}
\sigma ^{2}(k,h,j)\longmapsto Vol\left[ \sigma _{hyp}^{3}(\infty ,k,h,j)%
\right] = \\
=\mathcal{L}(\theta _{j\,\,kh})+\mathcal{L}(\theta _{khj})+\mathcal{L}%
(\theta _{hj\,k}),  \notag
\end{gather}
which is well-defined since, due to the symmetries of the dihedral angles
of $\sigma _{hyp}^{3}(\infty ,k,h,j)$, \ the valutation of $Vol\left[ \sigma
_{hyp}^{3}(\infty ,k,h,j)\right] $\ is independent from which vertex of the
tetrahedron is actually mapped to $\infty $, (see \cite{benedetti}, prop. C.2.8).
Thus, we can compute the volume of the three-dimensional hyperbolic manifold 
$V_{hyp}\setminus K$ as 
\begin{gather}
Vol\,\left[ V_{hyp}\left( \left\{ \Theta (k)\right\}
_{k=1}^{N_{0}(T)}\right)\setminus K \right] =\label{Hyppvolume}\\
\sum_{\left\{ \sigma ^{2}(k,h_{\alpha
},h_{\alpha +1})\right\} }^{N_{2}(T)}\left[ \mathcal{L}(\theta _{\alpha
+1\,,\,k,\,\alpha })+\mathcal{L}(\theta _{k,\,\alpha ,\,\alpha +1})+\mathcal{%
L}(\theta _{\alpha ,\,\alpha +1,\,k})\right] , \notag
\end{gather}
where the summation extends over all triangles $\sigma ^{2}(k,h_{\alpha
},h_{\alpha +1})$ in the Regge triangulated surface $|T_{l}|\rightarrow M$.
Equivalently, in terms of the complex moduli $\zeta _{\alpha
+1\,,\,k,\,\alpha }$ of the triangles $\sigma ^{2}(k,h_{\alpha },h_{\alpha
+1})$, we get
\begin{gather}
Vol\,\left[ V_{hyp}\left( \left\{ \Theta (k)\right\}
_{k=1}^{N_{0}(T)}\right)\setminus K \right] = \\
\sum_{\left\{ \sigma ^{2}(k,h_{\alpha },h_{\alpha +1})\right\} }^{N_{2}(T)}
\left[ \mathcal{L}(\arg \zeta _{\alpha +1\,,\,k,\,\alpha })+\mathcal{L}(\arg
\zeta _{k,\,\alpha ,\,\alpha +1})+\mathcal{L}(\arg \zeta _{\alpha ,\,\alpha
+1,\,k})\right] .  \notag
\end{gather}
It is worthwhile to remark that if one computes the Hessian of $Vol\,\left[
V_{hyp} \right] $
with respect the angular variables $\left\{ \theta _{\alpha
+1\,,\,k,\,\alpha }\right\} $ of the generic triangle $\sigma
^{2}(k,h_{\alpha },h_{\alpha +1})$ one \ gets 
\begin{gather}
H_{k,\,k}\doteq \frac{\partial ^{2}}{\partial \left. \theta _{\alpha
+1\,,\,k,\,\alpha }\right. ^{2}}\,Vol\left[ V_{hyp}\left( \left\{ \Theta
(k)\right\} _{k=1}^{N_{0}(T)}\right) \right] =-\cot \,\theta _{\alpha
+1\,,\,k,\,\alpha }= \label{formH} \\
=\frac{l^{2}(h_{\alpha +1},k)+l^{2}(k,h_{\alpha })-l^{2}(h_{\alpha
},h_{\alpha +1})}{4\,\Delta (\alpha +1\,,\,k,\,\alpha )}\,,  \notag \\
H_{\alpha ,\,\alpha }\doteq \frac{\partial ^{2}}{\partial \left. \theta
_{k,\,\alpha ,\,\alpha +1}\right. ^{2}}\,Vol\left[ V_{hyp}\left( \left\{
\Theta (k)\right\} _{k=1}^{N_{0}(T)}\right) \right] =-\cot \,\theta
_{k,\,\alpha ,\,\alpha +1}=  \notag \\
=\frac{l^{2}(k,h_{\alpha })+l^{2}(h_{\alpha },h_{\alpha +1})-l^{2}(h_{\alpha
+1},k)}{4\,\Delta (k,\alpha ,\,\,\alpha +1)}\,,  \notag \\
H_{\alpha +1,\,\alpha +1}\doteq \frac{\partial ^{2}}{\partial \left. \theta
_{\alpha ,\,\alpha +1,\,k}\right. ^{2}}\,Vol\left[ V_{hyp}\left( \left\{
\Theta (k)\right\} _{k=1}^{N_{0}(T)}\right) \right] =-\cot \,\theta _{\alpha
,\,\alpha +1,\,k}\,=  \notag \\
=\frac{l^{2}(h_{\alpha },h_{\alpha +1})+l^{2}(h_{\alpha
+1},k)-l^{2}(k,h_{\alpha })}{4\,\Delta (\alpha ,\,\,\alpha +1,k)},  \notag
\end{gather}
where $\Delta \doteq \Delta (\alpha +1\,,\,k,\,\alpha )$ denotes, up to cyclic
permutation, the Euclidean area of the triangle \ $\sigma ^{2}(k,h_{\alpha },h_{\alpha +1})$, (see paragraph \ref{megeom}).
From (\ref{formH}) we  get
\begin{gather}
l^{2}(h_{\alpha +1},k)=2\,\Delta (H_{\alpha +1,\,\alpha +1}+H_{k,\,k}), \\
l^{2}(k,h_{\alpha })=2\,\Delta \left( H_{k,\,k}+H_{\alpha ,\,\alpha }\right)
,  \notag \\
l^{2}(h_{\alpha },h_{\alpha +1})=2\,\Delta (H_{\alpha ,\,\alpha }+H_{\alpha
+1,\,\alpha +1}),  \notag
\end{gather}
which provide sign conditions on $H_{lm}$. Actually,
it is relatively easy (\cite{rivin1}) to show that the restriction of \ the Hessian of $Vol\,\left[
V_{hyp}\setminus K \right] $ to \ \ the local Euclidean
structure on each $\sigma ^{2}(k,h_{\alpha },h_{\alpha +1})$ \ is
negative-definite. This latter remark implies that (minus) the Hessian of the hyperbolic volume can be used as a natural quadratic form on the space of deformations of the Euclidean structures associated with random Regge triangulations and which naturally pairs with the Weil-Petersson measure (\ref{EucWP}) on moduli space. It is also clear that formally the hyperbolic volume (\ref{Hyppvolume}) does not require the existence of a complete hyperbolic structure on the support space $V\setminus K$, and we may well associate the function (\ref{Hyppvolume}) to $V\setminus K$. However, 
the existence of a complete hyperbolic structure implies that such a volume function is a topological invariant by Mostow rigidity. Moreover, one can formulate the so-called volume conjecture (R. Kashaev  and H. and J. Murakami) \cite{Kashaev95, Kashaev97, Murakami}, (and \cite{Ohtsuki} for a review),
which, in our setting, may be phrased by stating that if $K$ is not a split link and  $J_{n}(K;t)$ is its colored Jones polynomial
associated with the n-dimensional irreducible representation of $sl_{2}(%
\mathbb{C})$, then
\begin{equation}
2\pi \,\,\lim_{n\rightarrow \infty }\,\frac{\ln \,\left| J_{n}(K;\exp \left[ 
\frac{2\pi i}{n}\right] )\right| }{n}=Vol\,\left[ V_{hyp}\left( \left\{ \Theta (k)\right\}
_{k=1}^{N_{0}(T)}\right)\setminus K \right]
\label{Kash}
\end{equation}
(in the standard formulation of the volume conjecture the role of the support space 
\linebreak $V(\{\Theta (k)\}_{k=1}^{N_{0}(T)})$ is played by  $\mathbb{S}^{3}$, and one assumes that the complement  $\mathbb{S}^{3}\backslash K$ of the link $K$ admits a (complete)
hyperbolic structure). $J_{n}(K;t)$ is defined through the $n$-dimensional irreducible representations of the quantum group $U_{q}(sl(2,\mathbb{C}))$. For some hyperbolic knots in $\mathbb{S}^{3}$, in particular for the figure eight knot \cite{Hmurakami} (and for torus links, which are non-hyperbolic and yield $0$ on the right member of (\ref{Kash})), the conjecture has been proved, (see also \cite{Baseilhac} for a deep analysis). This connection between knot polynomials and hyperbolic volume has been actually promoted to be  part of a more general conjecture \cite{CS} relating the asymptotics of the colored Jones polynomials to the Chern-Simons invariant 
\begin{equation}
2\pi \,i\cdot \lim_{n\rightarrow \infty }\frac{\ln \,J_{n}(K;\exp \left[ 
\frac{2\pi i}{n}\right] )}{n}=CS\,\left[ V_{hyp}\diagup K\right] +i\,Vol\,%
\left[ V_{hyp}\diagup K\right] 
\end{equation}
and
\begin{equation}
\lim_{n\rightarrow \infty }\frac{J_{n+1}(K;\exp \left[ \frac{2\pi i}{n}%
\right] )}{J_{n}(K;\exp \left[ \frac{2\pi i}{n}\right] )}=\exp \left( \frac{1%
}{2\pi i}\left( CS\,\left[ V_{hyp}\diagup K\right] +i\,Vol\,\left[
V_{hyp}\diagup K\right] \right) \right) 
\end{equation}
where again we have formally referred all quantities to $V_{hyp}\diagup K$,
in particular $CS\,\left[ V_{hyp}\diagup K\right] $ is the Chern-Simons
invariant of the connection defined by the hyperbolic metric on $%
V_{hyp}\diagup K$. It should be clear that these statements have a status quite more conjectural then the original ones owing to the conical nature of $V_{hyp}\diagup K$, nonetheless they are reasonable in view of the holographic principle. Recall that 
a geometrical version of \emph{classical} holography is familiar in hyperbolic geometry as the Ahlfors-Bers theorem which applies to hyperbolic manifolds $V$ containing a compact subset determining a conformal structure on the boundary at $\infty$ of $V$. In such a case the geometry of $V$ is uniquely determined by such induced conformal structure at $\infty $. It should be clear from its very set-up that our approach to closed/open duality is, geometricaly speaking, holographic in nature. Roughly speaking it is akin to a simplicial version of Ahlfors-Bers theorem, (for a serious analysis of this issue for conical hyperbolic manifolds see  
\cite{Moroianu2}). At this stage it is important to refer to the remarkable paper \cite{krasnov} which examines the connection between the
volume of hyperbolic manifolds, the AdS/CFT correspondence and moduli space geometry. It would certainly interesting to analyze in depth the possible relation between their approach and our framework.


\newpage


\thebibliography{100}
\bibitem{Gop-Vafa}
Gopakumar, R., and Vafa, C., 
\emph{Adv. Theor. Math. Phys.} {\bf 3}, 1415
 (1999), [hep-th/9811131].

\bibitem{Gop1}
  R.~Gopakumar,
  \emph{``From free fields to AdS,''}
  Phys.\ Rev.\ D {\bf 70} (2004) 025009
  [arXiv:hep-th/0308184].

\bibitem{Gop2}
  R.~Gopakumar,
  \emph{``From Free Fields To Ads. II,''}
  Phys.\ Rev.\ D {\bf 70} (2004) 025010
  [arXiv:hep-th/0402063].

\bibitem{Gopproc}
  R.~Gopakumar,
 \emph{``Free field theory as a string theory?,''}
  Comptes Rendus Physique {\bf 5} (2004) 1111
  [arXiv:hep-th/0409233].

\bibitem{Gop3}
  R.~Gopakumar,
  \emph{``From free fields to AdS. III,''}
  Phys.\ Rev.\ D {\bf 72} (2005) 066008
  [arXiv:hep-th/0504229]
 
\bibitem{Gop4}
  J.~R.~David and R.~Gopakumar,
  \emph{``From spacetime to worldsheet: Four point correlators,''}
  arXiv:hep-th/0606078.

\bibitem{Gai-Ras}
  D.~Gaiotto and L.~Rastelli,
  \emph{``A paradigm of open/closed duality: Liouville D-branes and the  Kontsevich
  model''}
  JHEP {\bf 0507} (2005) 053
  [arXiv:hep-th/0312196].

\bibitem{Aharony}
O.~Aharony, Z.~Komargodski and S.~S.~Razamat,
\emph{``On the worldsheet theories of strings dual to free large N gauge theories''},
JHEP {\bf 0605} (2006) 16,
[arXiv:hep-th/06020226].

\bibitem{Akhmedov}
  E.~T.~Akhmedov,
  \emph{``Expansion in Feynman graphs as simplicial string theory,''}
  JETP Lett.\  {\bf 80} (2004) 218
  [Pisma Zh.\ Eksp.\ Teor.\ Fiz.\  {\bf 80} (2004) 247]
  [arXiv:hep-th/0407018].

\bibitem{strebel} K. Strebel, 
\emph{Quadratic differentials}, Springer-Verlag (1984).

\bibitem{mulase} M. Mulase, M. Penkava,
\emph{``Ribbon graphs, quadratic differentials on
Riemann surfaces, and algebraic curves defined over $\overline{\mathbb{Q}}$''},
The Asian Journal of Mathematics {\bf 2}, 875-920 (1998)  
[math-ph/9811024 v2].

\bibitem{Penner1} R. C. Penner 
\emph{The decorated Teichm\"{u}ller space of punctured surfaces,} 
Comm.\ Math.\ Phys.\ {\bf 113} (1987), 299-339,

\bibitem{ThurstonB} W. P. Thurston, 
\emph{Three-Dimensional Geometry and Topology} 
Vol.I (edited by S. Levi) Princeton Univ. Press (1997).

\bibitem{Kaufmann} R. Kaufmann, R.C. Penner, 
\emph{Closed/open string diagrammatics}, 
arXiv:math.GT/0603485.

\bibitem{mirzakhani1} M. Mirzakhani, 
\emph{Simple geodesics and Weil-Petersson volumes of moduli spaces of bordered Riemann surfaces}, Preprint (2005).

\bibitem{mirzakhani2} M. Mirzakhani, 
\emph{Weil-Petersson volumes and intersection theory on the moduli spaces of curves}, 
Preprint (2005).

\bibitem{Kashaev95} R. M. Kashaev  \emph{``A link invariant from quantum dilogarithm''}, Modern Phys. Lett. A, {\bf 10} (1995), 1409-1418.

\bibitem{Kashaev97} R. M. Kashaev  \emph{``The hyperbolic volume of knots from the quantum dilogarithm''}, Lett. Math. Phys. {\bf 39} (1997), 269-275.

\bibitem{Murakami} H. Murakami and J. Murakami  
\emph{``The colored Jones polynomials and the simplicial volume of a knot''}, 
Acta Math. {\bf 186} (2001), 85-104.

\bibitem{Hmurakami} H. Murakami, \emph{``Kashaev's invariant and the volume of a hyperbolic knot after Y. Yokota''} from: Physics and combinatorics 1999 (Nagoya)", World Sci.,
River Edge, NJ (2001) 244.

\bibitem{CS} H. Murakami, J. Murakami, M. Okamoto, T. Takata, and Y. Yokota 
\emph{Kashaev's Conjecture and the Chern-Simons Invariants of Knots and Links} 
Experimental Mathematics {\bf 11} (2002) 427.

\bibitem{Mondello} G.~Mondello \emph{``Triangulated Riemann surfaces
with boundary and the Weil-Petersson Poisson structure''},
math.DG/0610698.

\bibitem{regge}
T.~Regge and R.~M.~Williams,
\emph{``Discrete Structures In Gravity,''}
J.\ Math.\ Phys.\  {\bf 41} (2000) 3964
[arXiv:gr-qc/0012035],\\
H.~Hamber \emph{``Simplicial quantum gravity''}, lecture notes published
in \emph{Gauge theories, critical phenomena and random systems},
Proceeding of the 1984 Les Houches Summer School, Session XLIII (edited
by K.~Osterwalder and R.~Stora, North Holland, 1986).

\bibitem{carfora1} J. Ambj\o rn, M. Carfora, A. Marzuoli, \emph{The geometry of dynamical
triangulations}, Lecture Notes in Physics {\bf m50}, Springer Verlag (1997).

\bibitem{ambjorn} J. Ambj\"orn, B. Durhuus, T. Jonsson, 
\emph{``Quantum Geometry''},
Cambridge Monograph on Mathematical Physics, Cambridge Univ. Press
(1997).

\bibitem{carfora} M. Carfora, A. Marzuoli, 
\emph{``Conformal modes in simplicial quantum gravity and the Weil-Petersson volume of moduli space''},  
Adv.Math.Theor.Phys. {\bf 6}(2002) 357
[arXiv:math-ph/0107028].

\bibitem{rivin1} I. Rivin 
\emph{``Euclidean Structures on Simplicial Surfaces and Hyperbolic Volume''} 
Ann.\ of Math. {\bf 139} (1994) 553.

\bibitem{benedetti} R. Benedetti, C. Petronio 
\emph{``Lectures on Hyperbolic Geometry,''} 
Springer-Verlag, New York (1992).

\bibitem{Baird} P. Baird, 
\emph{``Riemannian twistors and Hermitian structures on low-dimensional space forms''}, 
Jour. of Math. Phys. {\bf 33}  (1992) 3340. 

\bibitem{Penner2} R. C. Penner 
\emph{``Weil-Petersson volumes''} 
Jour.\ Diff.\ Geom.\ {\bf 35} (1992) 559. 

\bibitem{witten} E. Witten, 
\emph{``Two dimensional gravity and intersection theory on moduli space''}, 
Surveys in Diff. Geom. {\bf 1.} (1991) 243.

\bibitem{schumacher} G. Schumacher, S. Trapani, 
\emph{``Estimates of Weil-Petersson volumes via effective divisors''}, 
Commun. Math. Phys. {\bf222}, (2001) 1
[arXiv:math.AG/0005094 v2].

\bibitem{grushevsky} 
S. Grushevsky, 
\emph{``Explicit upper bound for the Weil-Petersson volumes''} 
Math. Ann. {\bf 321}, (2001) 1
[arXiv:math.AG/0003217].

\bibitem{zograf} P. Zograf, 
\emph{``Weil-Petersson volumes of moduli spaces
of curves and curves and the genus expansion in two dimensional''} 
[arXiv:math.AG/9811026]

\bibitem{manin}Y. I. Manin, P. Zograf, 
\emph{``Invertible cohomological filed theories
and Weil-Petersson volumes''} 
Annales de l' Institute Fourier, {\bf Vol. 50}, (2000) 519
[arXiv:math-ag/9902051].

\bibitem{kontsevich} M. Kontsevich, 
\emph{``Intersection theory on moduli space of curves''},
Commun. Math. Phys. {\bf147}, (1992) 1.

\bibitem{looijenga} E. Looijenga, 
\emph{``Intersection theory on Deligne-Mumford compactifications''}, 
S\'{e}minaire Bourbaki, (1992-93), 768.

\bibitem{Moore} S. Cordes, G. Moore and S. Ramgoolan, 
\emph{``Lectures on 2D Yang-Mills Theory, Equivariant Cohomology and Topological Field Theories''}, 
Nucl.\ Phys.\ Proc.\ Suppl.\  {\bf 41} (1995) 184
[arXiv:hep-th/9411210].

\bibitem{PetronioPorti}  C. Petronio and J. Porti 
\emph{``Negatively Oriented Ideal Triangulations and a Proof of Thuston's Hyperbolic Dehn Filling Theorem''}, 
Expo. Math. {\bf 18} (2001) 1,
[arXiv:math.GT/9901045].

\bibitem{PetronioFrigerio} F. Costantino, R. Frigerio, B. Martelli and C. Petronio 
\emph{``Triangulations of 3-manifolds, hyperbolic relative handlebodies, and Dehn filling''},
arXiv:math.GT/0402339.

\bibitem{Moroianu} S. Moroianu, J-M. Schlenker 
\emph{``Hyperbolic Cone-Manifolds with Singular infinity''}, 
Preprint (2006).

\bibitem{Ohtsuki}  T. Ohtsuki (ed) 
\emph{``Problems on invariants of knots and 3-manifolds''} 
in:  T. Kohno, T. Le, J. Murakami, J. Roberts and V Turaev (eds.), 
\emph{``Invariants of knots and 3-manifolds''} 
Geom. and Topology Monographs, Vol. {\bf 4} (2002) 377.

\bibitem{Baseilhac}  S. Baseilhac and R. Benedetti 
\emph{``QHI, 3-manifolds scissors congruence classes and the volume conjecture''}
 in: \emph{``Invariants of knots and 3-manifolds''}, T. Ohtsuki et al. eds., 
Geom. and Topology Monographs, Vol. {\bf 4} (2002) 13-28. 
[arXiv:math.GT/0211053].

\bibitem{Moroianu2} S. Moroianu, J-M. Schlenker 
\emph{``Quasi-Fuchsian manifolds with particles''} 
arXiv:math.DG/0603441.

\bibitem{krasnov} K. Krasnov and J.-M. Schlenker  
\emph{``On the renormalized volume of hyperbolic 3-manifolds''}  
arXiv:math.DG/0607081.
\end{document}